\documentclass[twocolumn]{aastex631}

\usepackage{lineno}

\begin{document}

\title{Non-Detections of Helium in the Young Sub-Jovian Planets K2-100b, HD 63433b, \& V1298\,Tau\,c}%: Evidence for Magnetic Field Effects \& Confinement}

\correspondingauthor{Munazza K. Alam \& James Kirk}
\email{malam@stsci.edu, j.kirk22@imperial.ac.uk}

\author[0000-0003-4157-832X]{Munazza K. Alam}
\altaffiliation{These authors contributed equally to this work.}
\affiliation{Space Telescope Science Institute, 3700 San Martin Drive, Baltimore, MD 21218, USA}
\affiliation{Carnegie Earth \& Planets Laboratory, 5241 Broad Branch Road NW, Washington, DC 20015, USA}

\author[0000-0002-4207-6615]{James Kirk}
\altaffiliation{These authors contributed equally to this work.}
\affiliation{Department of Physics, Imperial College London, Prince Consort Road, London, SW7 2AZ, UK}
\affiliation{Center for Astrophysics $|$ Harvard \& Smithsonian, 60 Garden Street, Cambridge, MA 02138, USA}

\author[0000-0002-2248-3838]{Leonardo A. Dos Santos}
\affiliation{Space Telescope Science Institute, 3700 San Martin Drive, Baltimore, MD 21218, USA}

\author[0000-0003-0473-6931]{Patrick McCreery}
\affiliation{William H. Miller III Department of Physics and Astronomy, Johns Hopkins University, Baltimore, MD 21218, USA}

\author[0000-0002-3900-5111]{Andrew P. Allan}
\affiliation{Leiden Observatory, Leiden University, P.O. Box 9513, 2300 RA Leiden, The Netherlands}

\author[0000-0002-4856-7837]{James E. Owen}
\affiliation{Department of Physics, Imperial College London, Prince Consort Road, London, SW7 2AZ, UK}

\author[0000-0001-5371-2675]{Aline A. Vidotto}
\affiliation{Leiden Observatory, Leiden University, P.O. Box 9513, 2300 RA Leiden, The Netherlands}

\author[0000-0002-1199-9759]{Romain Allart}
\affiliation{D\'epartement de Physique, Institut Trottier de Recherche sur les Exoplan\`etes, Universit\'e de Montr\'eal, Montr\'eal, Qu\'ebec, H3T 1J4, Canada
\thanks{Trottier Postdoctoral Fellow}}

\author[0000-0002-9148-034X]{Vincent Bourrier}
\affiliation{Observatoire Astronomique de l'Universit\'e de Gen\`eve, Chemin Pegasi 51b, CH-1290 Versoix, Switzerland}

\author[0000-0001-9513-1449]{N\'estor Espinoza}
\affiliation{Space Telescope Science Institute, 3700 San Martin Drive, Baltimore, MD 21218, USA}

\author[0000-0002-3641-6636]{George W. King}
\affiliation{Department of Astronomy, University of Michigan, Ann Arbor, MI 48109, USA}

\author[0000-0003-3204-8183]{Mercedes L\'opez-Morales} 
\affiliation{Center for Astrophysics $|$ Harvard \& Smithsonian, 60 Garden Street, Cambridge, MA 02138, USA}

\author[0000-0002-7990-9596]{Julia V. Seidel}
\affiliation{European Southern Observatory,  Alonso de C\'ordova 3107, Vitacura, Regi\'on Metropolitana, Chile}

\shortauthors{Alam \& Kirk et al.}
 
\begin{abstract}
% We search for excess in-transit absorption of neutral helium at 1.083 $\mu$m in the atmospheres of the young ($<$800\,Myr) sub-Jovian (0.2--0.5\,$\rm R_{J}$) planets HD 63433b, K2-100b, and V1298\,Tau\,c using high-resolution (R$\sim$25,000) transit observations taken with Keck II/NIRSPEC. Our observations do not show evidence of helium absorption for any of the planets in our sample. Based on comparisons to one-dimensional Parker wind models as well as one-dimensional hydrodynamic escape models, we find \textbf{3$\sigma$} upper limit\textbf{s} on the excess helium absorption of $\sim$0.47--1.13\% (EW \textbf{$<$} 2.52--8.49\,m\AA) for the planets in our sample. We also consider the effects of magnetic fields and stellar wind confinement on the He line, and find that both of these mechanisms predict upper limits that are consistent with our non-detections. Our measured upper limits on the helium absorption for these planets are consistent with predicted trends in system age and He equivalent width from 1D hydrodynamic models.

We search for excess in-transit absorption of neutral helium at 1.083 $\mu$m in the atmospheres of the young ($<$800\,Myr) sub-Jovian (0.2--0.5\,$\rm R_{J}$) planets HD 63433b, K2-100b, and V1298\,Tau\,c using high-resolution (R$\sim$25,000) transit observations taken with Keck II/NIRSPEC. Our observations do not show evidence of helium absorption for any of the planets in our sample. We calculate $3\sigma$ upper limits on the planets' excess helium absorption of $<$0.47\% for HD 63433b, $<$0.56\% for K2-100b and $<$1.13\% for V1298 Tau c. In terms of equivalent width, we constrain these to $<$2.52, $<$4.44 and $<$8.49\,m\AA\ for HD 63433b, K2-100b and V1298 Tau c, respectively. We fit our transmission spectra with one-dimensional Parker wind models to determine upper limits on the planets' mass-loss rates of $<$7.9$\times10^{10}$, $<$1.25$\times10^{11}$ and $<$7.9$\times10^{11}$\,g\,s$^{-1}$. Our non-detections align with expectations from one-dimensional hydrodynamic escape models, magnetic fields, and stellar wind confinement. The upper limits we measure for these planets are consistent with predicted trends in system age and He equivalent width from 1D hydrodynamic models.

\end{abstract}


%%%%%%%%%%%%%%%%%%%%%%%%%%%%%%%%%%%%%%%%%%%%%%%%%%%%%%%%%
\section{Introduction} 
\label{sec:intro}
%%%%%%%%%%%%%%%%%%%%%%%%%%%%%%%%%%%%%%%%%%%%%%%%%%%%%%%%%

Short-period planets intermediate in size between Earth and Neptune (1--4\,\rm $R_{\oplus}$), are abundant in the galaxy (e.g., \citealt{Batalha13,Dressing15,Morton16}), but their formation and evolutionary histories are currently unknown. While those larger than $\rm 1.6\,R_{\oplus}$ likely possess large H/He envelopes comprising $\sim$1\% of their total mass, planets below this threshold have densities consistent with rocky compositions (e.g., \citealt{Rogers15}). The observed distribution of transiting planetary radii from \textit{Kepler} further reveals a dearth of exoplanets with radii between 1.5--2.0\,$\rm R_{\oplus}$ \citep{Fulton17}. While planet formation and core-powered evolution models can partly explain these results (e.g., \citealt{Ginzburg18,Loyd18}), photoevaporation -- the escape of planetary atmospheres caused by high-energy irradiation from their host stars -- may play an important role in sculpting these observed trends (e.g., \citealt{Lopez14,Owen17}).

Intermediate-sized planets are the best targets for probing atmospheric escape that changes the bulk characteristics of exoplanets. Jupiter-sized planets are too massive to lose a significant part of their primordial atmospheres during their evolution (e.g., \citealt{Lecavelier07}), whereas small Earth-sized planets lose their volatile-rich atmospheres too quickly (e.g., \citealt{Owen17}) and are not accessible with currently-available instruments (e.g., \citealt{Bourrier2017, Waalkes2019}). Some of the most extreme cases of ongoing atmospheric escape and evolution have been observed for planets with sizes similar to Neptune: GJ\,436b \citep{Ehrenreich15}, GJ\,3470b \citep{Bourrier18}, HAT-P-11b \citep{Allart18,Mansfield18,BJaffel2022}, and tentatively K2-18b \citep{DosSantos20}. Signatures of atmospheric escape imprint large ($\sim$0.1--5\% for He; $\gtrsim$10\% for H) signals on transit spectra that are readily detectable with current instruments.

% Given the abundance and unknown origins of sub-Neptunes and the successful detections of mass-loss around Neptunes, we have conducted a survey spanning both size regimes. Furthermore, while photoevaporation is predicted to be most significant at the earliest ages ($<$500 Myr; \citealt{Lopez12,Owen13}), there are few observational tests of this theory and a lack of planets sampling the upper limit of this theoretical age limit. Therefore, our sample also well-samples ages where photoevaporation should be dominant (23 Myr) up to ages older than we would expect photoevaporation to be dominant (750 Myr), enabling important tests of this theory. 

The majority of observations of atmospheric escape and extended atmospheres prior to 2018 were obtained with Lyman-$\alpha$ transmission spectroscopy using \textit{Hubble} (e.g., \citealt{Vidal-Madjar03,Lecavelier10,Ehrenreich15,Bourrier18,DosSantos20}). The stellar Lyman-$\alpha$ emission line is efficiently absorbed by the interstellar medium (ISM), limiting the detection of atmospheric escape to nearby ($\lesssim$\,60\,pc) targets. \citet{Spake18}, however, demonstrated that the metastable He triplet at 1.083\,$\mu$m -- which is devoid of ISM absorption -- is a viable probe for atmospheric escape (\citealt{Seager00,Oklopcic18}). Several studies have since demonstrated the feasibility of ground-based transmission spectroscopy to study atmospheric escape via the 1.083\,$\mu$m He triplet (e.g., \citealt{Allart18,Nortmann18,Alonso-Floriano19,Kirk20,Zhang21}). \citet{Kirk20} showed for the first time the capability of Keck II/NIRSPEC to achieve high-precision, high-confidence (30$\sigma$) detections of excess He absorption.

Photoevaporation models predict that planets lose a significant amount of their primordial H/He atmospheres at early ages ($\tau$\,$<$\,100-500\,Myr) when high-energy irradiation from the host star is the strongest (e.g., \citealt{Jackson12,Owen13}). 
Detecting and measuring the atmospheric escape rates from young planets can allow us to place more direct observational constraints on planetary evolution models at early ages. These young targets are further favorable for helium observations because the intense stellar X-ray and extreme ultraviolet (XUV) irradiation levels in the first 100 Myr of a planet’s lifetime lead to a metastable He line that is more likely to be populated \citep{Oklopcic19, Allan23}. Young transiting planets thereby represent an unprecedented opportunity to probe atmospheric escape and evolution in small, irradiated planets.

In this work, we present high-resolution spectroscopic observations of He {\sc i} at 1.083 $\mu$m using Keck II/NIRSPEC in the young ($<$800\,Myr) sub-Jovian (0.2--0.5\,$\rm R_{J}$) planets K2-100b \citep{Livingston18a},  HD 63433b \citep{Mann20}, and V1298\,Tau\,c \citep{David19} to measure the persistence and amplitude of the helium feature in young planets. The system and orbital properties of our sample are included in Table \ref{tab:params}. While photoevaporation is predicted to be most significant at the earliest ages ($<$500 Myr; \citealt{Lopez12,Owen13}), there are few observational tests of this theory and a lack of planets sampling the upper limit of this theoretical age limit. Therefore, our sample spans ages where photoevaporation should be dominant (23 Myr) up to ages older than we would expect photoevaporation to be dominant (750 Myr), enabling important tests of this theory.

The paper is structured as follows. \S \ref{sec:obs_dr} details the observations and data reduction procedures, and \S \ref{sec:analysis} describes our time-series analysis to obtain the helium transmission spectra for the planets in our sample. We  interpret our observations considering the effects of magnetic fields, stellar winds, and 1D hydrodynamic escape models in \S \ref{sec:interpretation}. In \S \ref{sec:discussion}, we contextualize our results in light of previous observations of these targets as well as other helium observations for both young and older exoplanets. We summarize our conclusions in \S \ref{sec:summary}.

% \documentclass{aastex}
% \begin{document}
\begin{deluxetable*}{lccc}
% \tabletypesize{\scriptsize}
% \tabletypesize{\tiny}
\tablewidth{0pt}
\tablecolumns{4}
\tablecaption{System Parameters for HD 63433b, V1298\,Tau\,c, \& K2-100\,b. Integrated fluxes received by each planet over the X-ray [0.517-12.4 nm], hard-EUV [10-36 nm], soft-EUV [36-92 nm] and mid-UV [91.2-320 nm] bands from the \citet{Allan23} model. Stellar, planetary, and system parameters are taken from (a) \citet{Mann20}, (b) \citet{Capistrant24}, (c) \citet{Jones15}, (d) Gaia DR2, (e) \citet{Barragan19}, (f) \citet{Brandt15}, (g) \citet{David19}, (h) \citet{Gaidos22}, (i) derived from mass-radius relation \citep{Chen2017}, (j) \citet{Stassun2019}. \label{tab:params}}
\tablehead{\colhead{}  & \colhead{HD 63433b}  & \colhead{V1298\,Tau\,c}  & \colhead{K2-100\,b} }
\startdata
\hline 
% \vspace{-1cm} \\
\sidehead{Stellar Parameters}  
\hline
% \vspace{-0.5cm}\\\hline  \vspace{-0.3cm} \\
Mass, $M_{\star}$ [$\rm M_{\odot}$]   	 & 0.99$\pm$0.03\tablenotemark{a} & 1.10$\pm$0.05\tablenotemark{g}   & 1.15$\pm$0.05\tablenotemark{e}  	\\
Radius, $R_{\star}$ [$\rm R_{\odot}$] 	 & 0.91$\pm$0.03\tablenotemark{a}	 & 1.34$\pm$0.05\tablenotemark{g}    & 1.24$\pm$0.05\tablenotemark{e}  	\\
Effective temperature, $T_{\mathrm{eff}}$ [K]  	 & 5640$\pm$74\tablenotemark{a}	 & $4962^{+88}_{-45}$\tablenotemark{d}   & 5945$\pm$110\tablenotemark{e}  	\\
Metallicity, [Fe/H]   & 0.03$\pm$0.05\tablenotemark{b} 	 & 0.139$\pm$0.03\tablenotemark{j}  & 0.22$\pm$0.09\tablenotemark{e}  	\\
Surface gravity, log(g) [cgs]             & 4.52$\pm$0.05\tablenotemark{b}  	 & 4.25$\pm$0.03\tablenotemark{g}   & 4.33$\pm$0.10\tablenotemark{e}  	\\
% Semi-amplitude, $K$	[m/s]					 & 8.075		 & 0.00   & 10.6\textbf{$\pm$3.0}\tablenotemark{e}    \\
Age [Myr] & 414$\pm$23\tablenotemark{c} & 23$\pm$4\tablenotemark{g}  & 750$\pm$5\tablenotemark{f} \\
$ F_{\rm X-ray}$ [erg/s/cm$^2$] & 5.39$\times$10$^{3}$ &   1.84$\times$10$^{5}$ &  1.98$\times$10$^{4}$  \\
$  F_{\rm hEUV}$ [erg/s/cm$^2$] & 4.64$\times$10$^{3}$ &  3.75$\times$10$^{4}$  & 1.70$\times$10$^{4}$  \\
$ F_{\rm sEUV} $ [erg/s/cm$^2$] & 6.75$\times$10$^{3}$ & 1.73$\times$10$^{4}$  & 4.60$\times$10$^{4}$ \\
$ F_{\rm mid-UV}$ [erg/s/cm$^2$] & 5.33$\times$10$^{6}$   & 1.19$\times$10$^{6}$ & 3.21$\times$10$^{6}$ \\
\hline 
% \vspace{-1cm} \\
\sidehead{Planetary Parameters}    
\hline
%\vspace{-0.5cm}\\
% \hline  \vspace{-0.3cm} \\
Mass, $M_{p}$ [$\rm M_{J}$]   		 	 & 0.0166\tablenotemark{i} 	 & 0.0839\tablenotemark{i}    & 0.0686$\pm$0.02\tablenotemark{e} 	\\
Radius, $R_{p}$ [$\rm R_{J}$] 		 	 & 0.192$\pm$0.009\tablenotemark{a} 	 & 0.499$\pm$0.03\tablenotemark{g}     & 0.346$\pm$0.01\tablenotemark{e}  	\\
Equilibrium temperature, $T_{eq}$ [K] 	 & 968$\pm$36\tablenotemark{a} 	 & 981$\pm$31\tablenotemark{g}    & 1841$\pm$41\tablenotemark{e} 	\\
Surface gravity, log(g) [cgs]         	 & 3.05\tablenotemark{i} 	 & 2.92\tablenotemark{i}    & 3.15$\pm$0.13\tablenotemark{e}	\\
\hline 
% \vspace{-1cm} \\
\sidehead{System Parameters}  
\hline %\vspace{-0.5cm}\\
%  \vspace{-0.3cm} \\
Systemic velocity, $\gamma$ [km/s]		 & $-$16.31$\pm$0.20\tablenotemark{d}   	  & +14.644$\pm$0.136\tablenotemark{h}         & +34.78$\pm$0.64\tablenotemark{d}  	   	\\
Period, $P$ [days]                     	 & 7.10793$\pm$0.0004\tablenotemark{a} 	 	  & 8.24958$\pm$0.00072\tablenotemark{g}         & 1.6739035$\pm$0.0000004\tablenotemark{e}	\\
Inclination, $i$ [$^{\circ}$]  			 & 89.38$^{+0.43}_{-0.64}$\tablenotemark{a}   	 	  & 88.49$^{+092}_{-0.72}$\tablenotemark{g}           & 81.27$\pm$0.37\tablenotemark{e}		 	\\
Scaled semi-major axis, $a/R_{\star}$  	 & 16.95$^{+0.34}_{-0.82}$\tablenotemark{a} 	 	  & 13.19$\pm$0.55\tablenotemark{g}    		& 5.21$\pm$0.13\tablenotemark{e}		 	\\
Radius ratio, $R_{p}/R_{\star}$ 	   		 & 0.02161$\pm$0.00055\tablenotemark{a}   	  & 0.0381$\pm$0.0017\tablenotemark{g}    		& 0.02867$\pm$0.00028\tablenotemark{e}	 	\\
Mid-transit time, $T_{0}$ [BJD]	    	 & $2458916.4526^{+0.0032}_{-0.0027}$\tablenotemark{a}  & $2457064.2797\pm0.0034$\tablenotemark{g}   & $2457140.71941\pm0.00027$\tablenotemark{e} \\
Transit duration [days]				 & 0.134$\pm$0.0014\tablenotemark{a}     	  & 0.194$\pm$0.005\tablenotemark{g} 		    & 0.067$\pm$0.0006\tablenotemark{e}     	    \\
 \enddata
\end{deluxetable*}
% \end{document}

%%%%%%%%%%%%%%%%%%%%%%%%%%%%%%%%%%%%%%%%%%%%%%%%%%%%%%%%%
\section{Observations \& Data Reduction} 
\label{sec:obs_dr}
%%%%%%%%%%%%%%%%%%%%%%%%%%%%%%%%%%%%%%%%%%%%%%%%%%%%%%%%%

%%%%%%%%%%%%%%%%%%%%%%%%%%%%%%%%%%%%%%%%%%%%%%%%%%%%%%%%%
\subsection{Observations}
\label{sec:obs}
%%%%%%%%%%%%%%%%%%%%%%%%%%%%%%%%%%%%%%%%%%%%%%%%%%%%%%%%%
We observed a single transit each of V1298\,Tau\,c, HD 63433b, and K2-100b with Keck II/NIRSPEC \citep{McLean98,Martin18} on UT 17 Dec 2020, UT 24 Dec 2020, and UT 08 Jan 2021, respectively. Our observations were taken as part of Program N165 (PI: Alam) using the NIRSPEC-1 filter, which covers the $Y$-band (0.947--1.121 $\mu$m; NIRSPEC orders 68-80) at a spectral resolution of R$\sim$25,000. Here we present the extracted NIRSPEC order 70 (1.0799--1.1014 $\mu$m) spectra only, which covers the metastable helium triplet at 1.0833 $\mu$m. 

To mitigate fringing effects, we did not use the `Thin' blocking filter (e.g., \citealt{Kasper20,Kirk22}). At the beginning and end of our transit observations, we obtained a set of 11 darks (including bias frames), 11 lamp flats, and two arcs (each composed of 10 co-added Ne, Ar, Xe, and Kr arc lamps). We obtained 28 spectra for V1298\,Tau\,c ($J$=8.6) with exposure times of 300 seconds each over 182 minutes, and acquired an additional 6 out-of-transit spectra for this target on UT 08 Jan 2021\footnote{\footnotesize{The first night of V1298\,Tau data was severely affected by bad weather, limiting the number of out-of-transit exposures to 3.}}. For HD 63433b ($J$=5.6), we acquired 372 spectra with exposure times of 30 seconds over the course of 303 minutes. For K2-100b ($J$=9.4), we obtained 26 600-second spectra over 255 minutes. For all of our observations, we used an ABBA nod pattern to remove sky background by A-B pair subtraction. We achieved an average SNR per pixel per exposure in order 70 of 199, 213, and 191 for V1298\,Tau\,c, HD 63433b, and K2-100b, respectively. %A summary of our observations is reported in Table \ref{tab:obs}.  %V1298 Tau c SNR/pix/exp: 213 (Dec) and 185 (Jan) -- report the average of the two instead?

%%%%%%%%%%%%%%%%%%%%%%%%%%%%%%%%%%%%%%%%%%%%%%%%%%%%%%%%%
\subsection{Data Reduction}
\label{sec:dr}
%%%%%%%%%%%%%%%%%%%%%%%%%%%%%%%%%%%%%%%%%%%%%%%%%%%%%%%%%
We reduced our observations using the {\tt REDSPEC}\footnote{\url{https://www2.keck.hawaii.edu/inst/nirspec/redspec.html}} software \citep{McLean03,McLean07}, which performs spatial rectification of tilted spectral orders on the detector, bad pixel interpolation, flat-fielding, and dark and bias subtraction. For the dark subtraction and flat-field correction steps in REDSPEC, we created a main dark and a main flat by median-combining our 22 darks and 22 flats, respectively. We used a second-order polynomial to correct for the tilt of the spectral orders on the detector. We then performed a wavelength calibration for order 70 using our arc lamp spectra and a second-order polynomial to map the measured arc line locations to the theoretical values. The spectra were then extracted in differenced A-B nod pairs to remove the sky background and OH emission lines, with an aperture width of 11 pixels.  

%%%%%%%%%%%%%%%%%%%%%%%%%%%%%%%%%%%%%%%%%%%%%%%%%%%%%%%%%
\subsection{Post-Processing}
\label{sec:post-processing}
%%%%%%%%%%%%%%%%%%%%%%%%%%%%%%%%%%%%%%%%%%%%%%%%%%%%%%%%%
We post-processed the extracted stellar spectra following the prescription outlined in \citet{Kirk22}, which we briefly summarize here. After extracting the wavelength-calibrated time-series spectra, we continuum-normalized the spectra using {\tt iSpec}\footnote{\url{https://www.blancocuaresma.com/s/iSpec}} \citep{Blanco-Cuaresma14,Blanco-Cuaresma19}. We fit cubic splines to the 1.080-1.095 $\mu$m portion of order 70, masking the helium triplet from our continuum calculation. 

\begin{figure}
    \centering
    \includegraphics[scale=0.55]{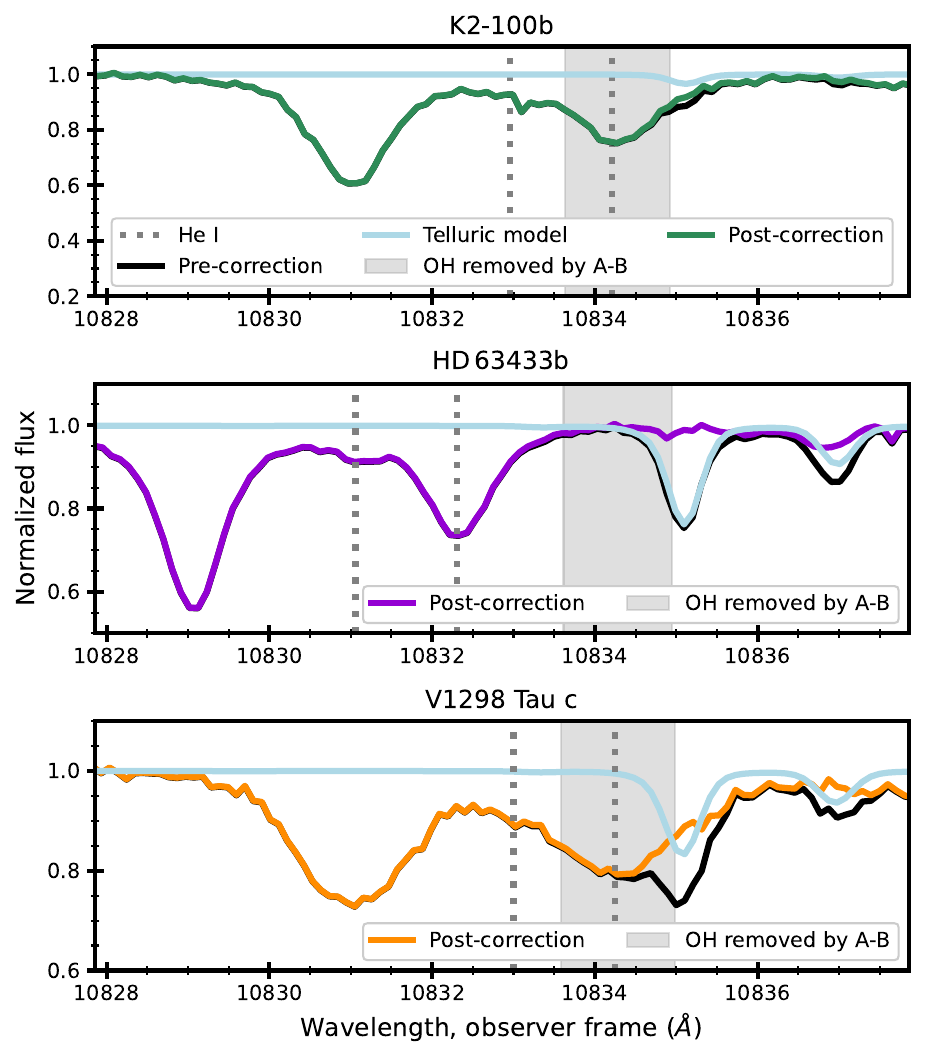}
    \caption{Examples of pre- (black) and post-telluric corrected spectra for K2-100b (green), HD 63433b (purple), and V1298\,Tau\,c (orange). The best-fitting \texttt{molecfit} H$_2$O models are shown in light blue. The telluric OH emission occurs in the shaded gray region, however, this feature is successfully removed by the A-B nod pair subtraction. The dashed gray lines indicate the location of the He triplet which are different in the observer frame due to the combination of the barycentric and systemic velocities.}
    \label{fig:molecfit_models}
\end{figure}

We then used {\tt molecfit}\footnote{\url{https://www.eso.org/sci/software/pipelines/skytools/molecfit}}  \citep{Kausch15,Smette15} to remove telluric features. We selected telluric absorption lines away from stellar absorption lines to constrain the {\tt molecfit} model, fitting for atmospheric H$_{2}$O only. For HD 63433b, we used seven telluric absorption lines and three for K2-100b. For V1298\,Tau\,c, we used four telluric absorption lines for the December observations and three for the January observations. We fitted the telluric absorption lines with a Gaussian with a FWHM initiated at 3.5 pixels. The FWHM of the Gaussian was fixed to 3.5 pixels for K2-100b following (\citealt{Zhang21,Kirk22}), due to difficulties in the ability of the data to constrain this parameter for this night. The fitted FWHMs of the telluric models for the other two targets were $3.33 \pm 0.16$ pixels (HD 63433b) and $3.73 \pm 0.26$ pixels (V1298\,Tau\,c). Example telluric-corrected spectra are shown in Figure \ref{fig:molecfit_models}.

To check our systematic errors in the wavelength solution near the He triplet, we cross-correlated our stellar spectra with model spectra that combined PHOENIX stellar atmosphere models \citep{Husser13} with a telluric model from the {\tt molecfit}\footnote{\url{https://www.eso.org/sci/software/pipelines/skytools/molecfit}} software \citep{Kausch15,Smette15}. By cross-correlating sub-sections of the spectral order, we found that the bluest wavelengths ($\leq 1.087$\,$\mu$m) were shifted from the truth by $\sim$30\,km\,s$^{-1}$ for K2-100b, $\sim$12\,km\,s$^{-1}$ for HD\,63433b and $\sim$-10\,km\,s$^{-1}$ for V1298\,Tau\,c. However, at the reddest wavelengths ($\geq 1.092$\,$\mu$m) the spectra were shifted by --2\,km\,s$^{-1}$, --5\,km\,s$^{-1}$ and 1\,km\,s$^{-1}$, respectively. We corrected for this wavelength distortion by fitting a quadratic polynomial to our wavelength-dependent cross-correlation results. As a final check to our wavelength solution, we fit a Voigt absorption profile to a non-saturated absorption line at 1.0842~$\mu$m in the same spectral order and measured the center of the fitted profile for each exposure. We then calculated its shift in Doppler velocity space from the average stellar spectrum, and correct the wavelength solution in the He triplet region based on this systematic Doppler velocity shift; this procedure was repeated for each exposure.

We then shifted all of the spectra from the observer rest frame to the stellar rest frame by correcting for the barycentric, systemic, and stellar reflex velocities \citep{Wright14}. To account for cosmic rays in our datasets, we constructed a median-combined spectrum and compared each spectral frame to the combined spectrum. We replaced data points that deviated by $>$4$\sigma$ from the median spectrum to the median of the surrounding pixels. We flagged 11 outliers for HD 63433b and 14 outliers for V1298 Tau c within $\pm$20 km/s of the helium triplet in the planets' rest frames but opted not to replace these to avoid removing real planetary signal. Due to the larger amplitude and greater number of outliers in the K2-100 data set (26 pixels within $\pm$20 km/s of the helium triplet, equal to 5\% of the total pixels) we replaced these to avoid anomalous signals caused by outliers. %after masking the spectra within $\pm$20 km/s of the helium triplet in the planet's rest frame to avoid removing real planetary signal. \textbf{From this exercise, we found that 11 data points were flagged for replacement for HD 63433b, 16 points for K2-100b, and 14 points for V1298 Tau c.}           

%%%%%%%%%%%%%%%%%%%%%%%%%%%%%%%%%%%%%%%%%%%%%%%%%%%%%%%%%
\section{Time-Series Analysis} 
\label{sec:analysis}
%%%%%%%%%%%%%%%%%%%%%%%%%%%%%%%%%%%%%%%%%%%%%%%%%%%%%%%%%

\begin{figure*}
    \centering
    \includegraphics[width=\textwidth]{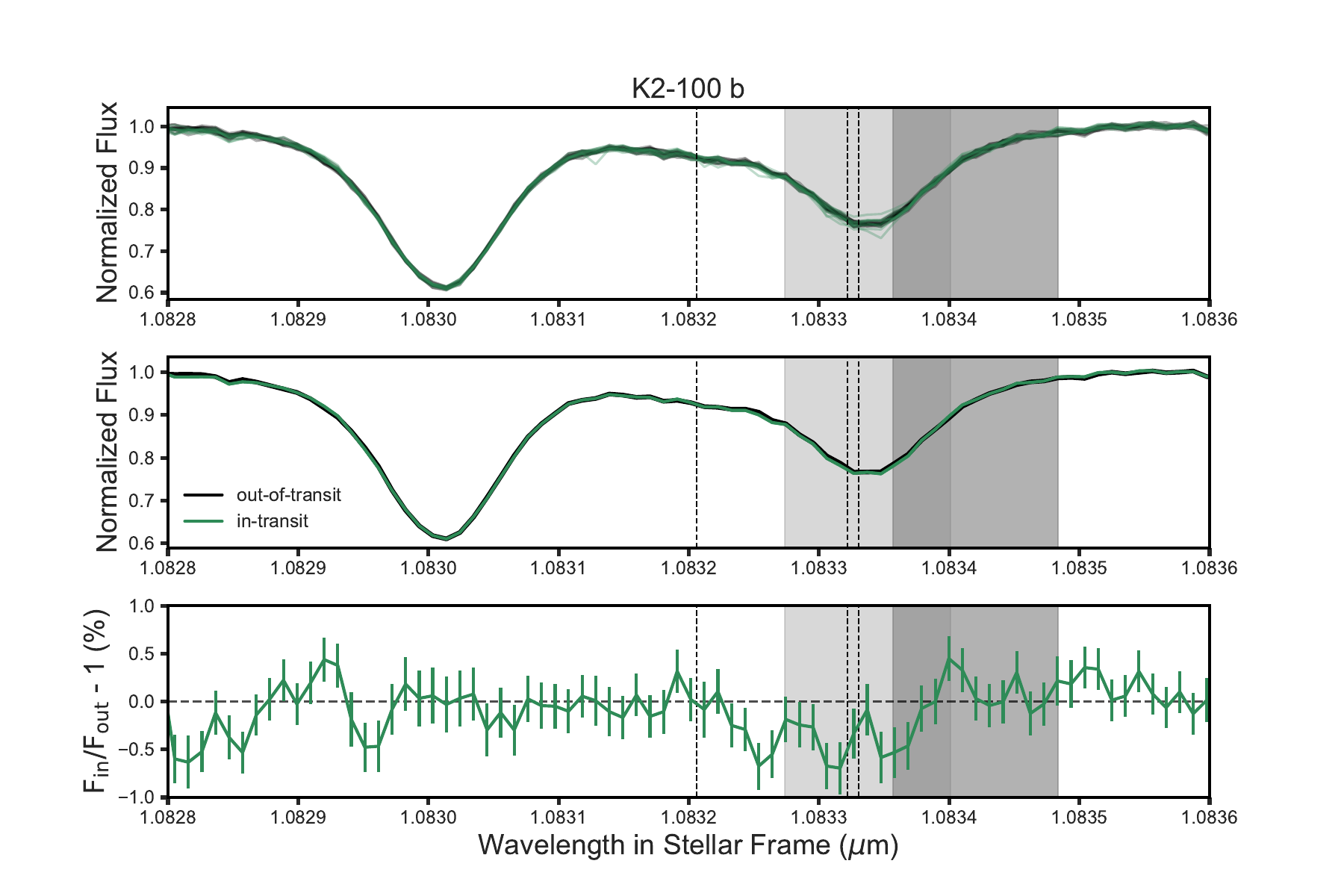}
    \caption{In-transit (green) and out-of-transit (black) stellar spectra of K2-100b for all individual spectral frames (top) and the mean spectra (middle), as well as the excess absorption (bottom) centered on the helium  triplet (dashed black lines). The OH emission (light gray) and H$_{2}$O absorption (dark gray) telluric regions are also indicated.}
    \label{fig:k2100_it_oot}
\end{figure*}

\begin{figure*}
    \centering
    \includegraphics[width=\textwidth]{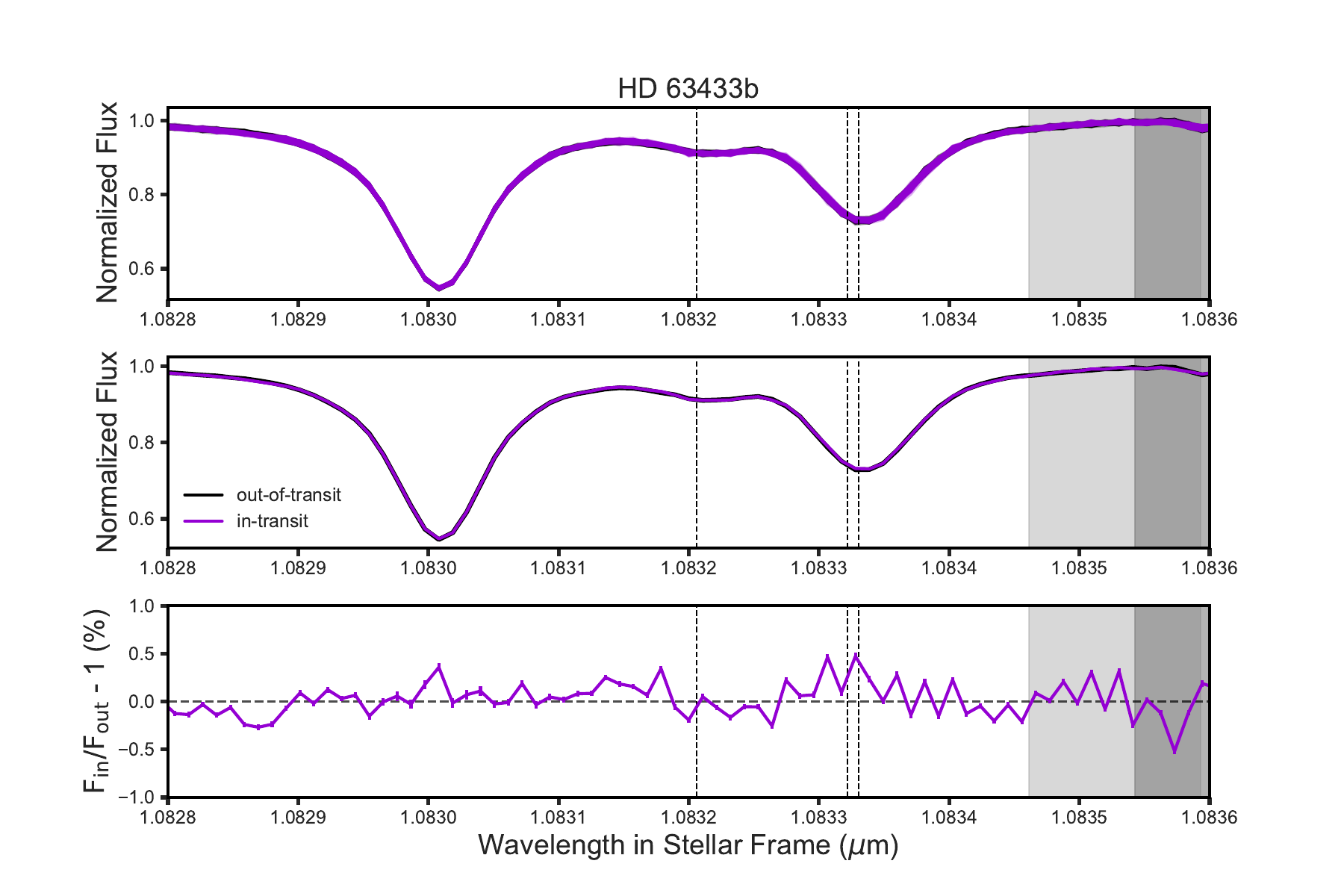}
    \caption{In-transit (purple) and out-of-transit (black) stellar spectra of HD 63433b for all spectral frames (top) and the mean spectra (middle), as well as the excess absorption (bottom) centered on the helium  triplet (dashed black lines). The OH emission (light gray) and H$_{2}$O absorption (dark gray) telluric regions are also indicated.}
    \label{fig:toi1726_it_oot}
\end{figure*}

\begin{figure*}
    \centering
    \includegraphics[width=\textwidth]{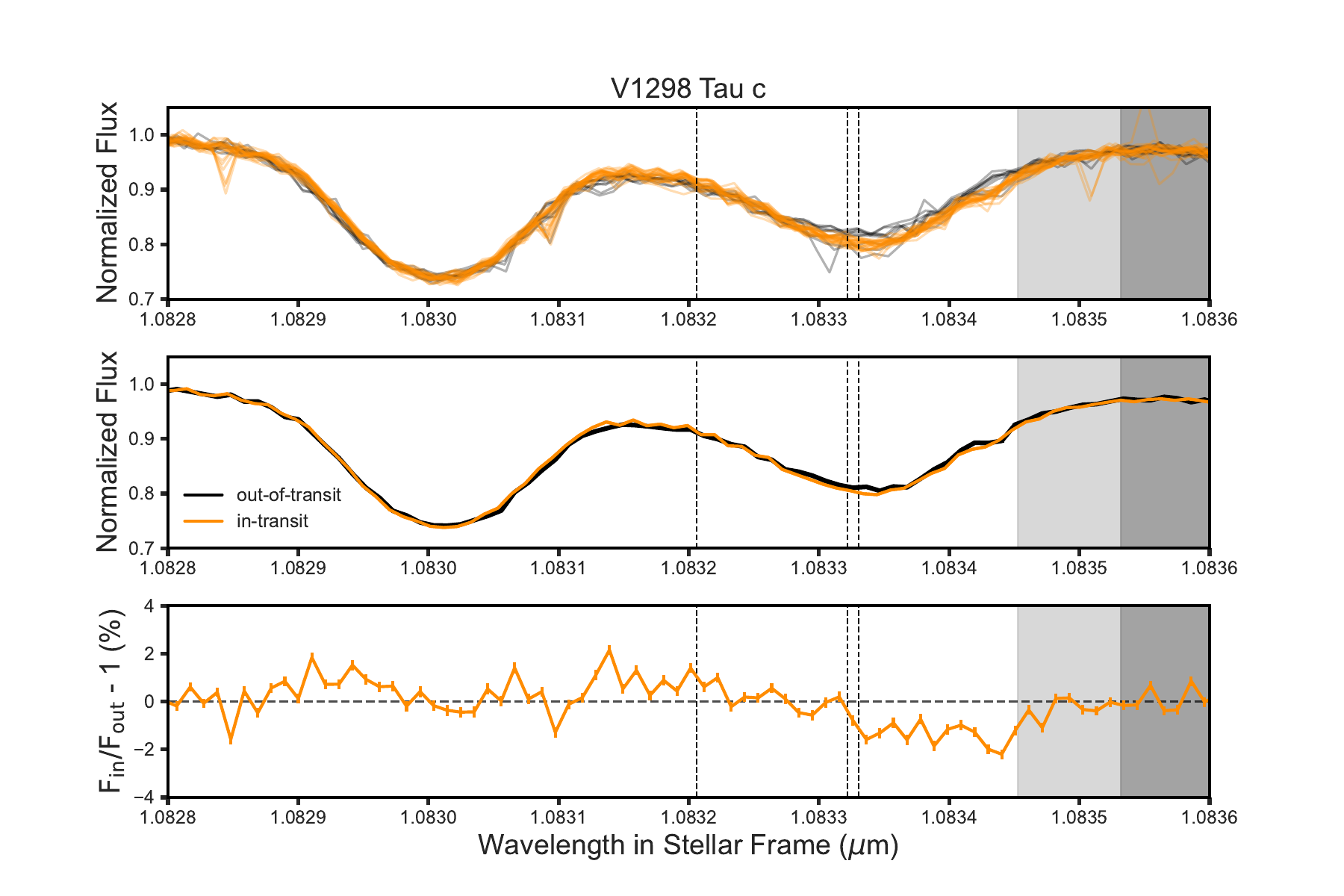}
    \caption{In-transit (orange) and out-of-transit (black) stellar spectra of V1298\,Tau\,c for all spectral frames (top) and the mean spectra (middle), as well as the excess absorption (bottom) centered on the helium  triplet (dashed black lines). The OH emission (light gray) and H$_{2}$O absorption (dark gray) telluric regions are also indicated.} %The color bar shows the S/N of our observations, where lighter colors indicate absorption and darker colors indicate emission.}
    \label{fig:v1298_it_oot}
\end{figure*}

\begin{figure*}
    \centering
    \includegraphics[width=\textwidth]{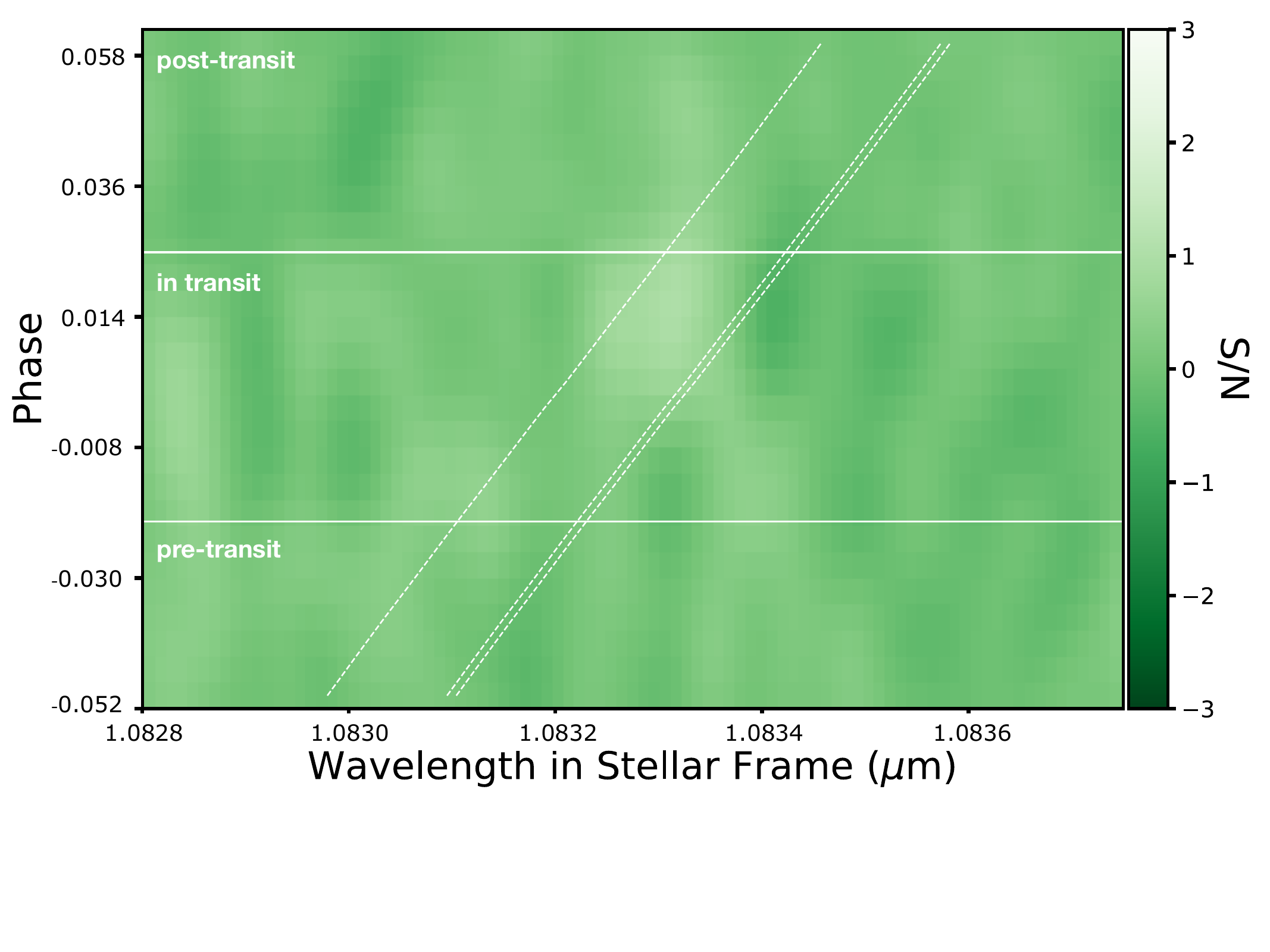}
    \caption{Phase-resolved signal-to-noise map for K2-100b centered on the helium triplet in the stellar rest frame, compared to the first and fourth contact points (solid white lines) and the planet's orbital motion (dashed white lines).}
    \label{fig:k2100_He_absorption_map}
\end{figure*}

\begin{figure*}
    \centering
    \includegraphics[width=\textwidth]{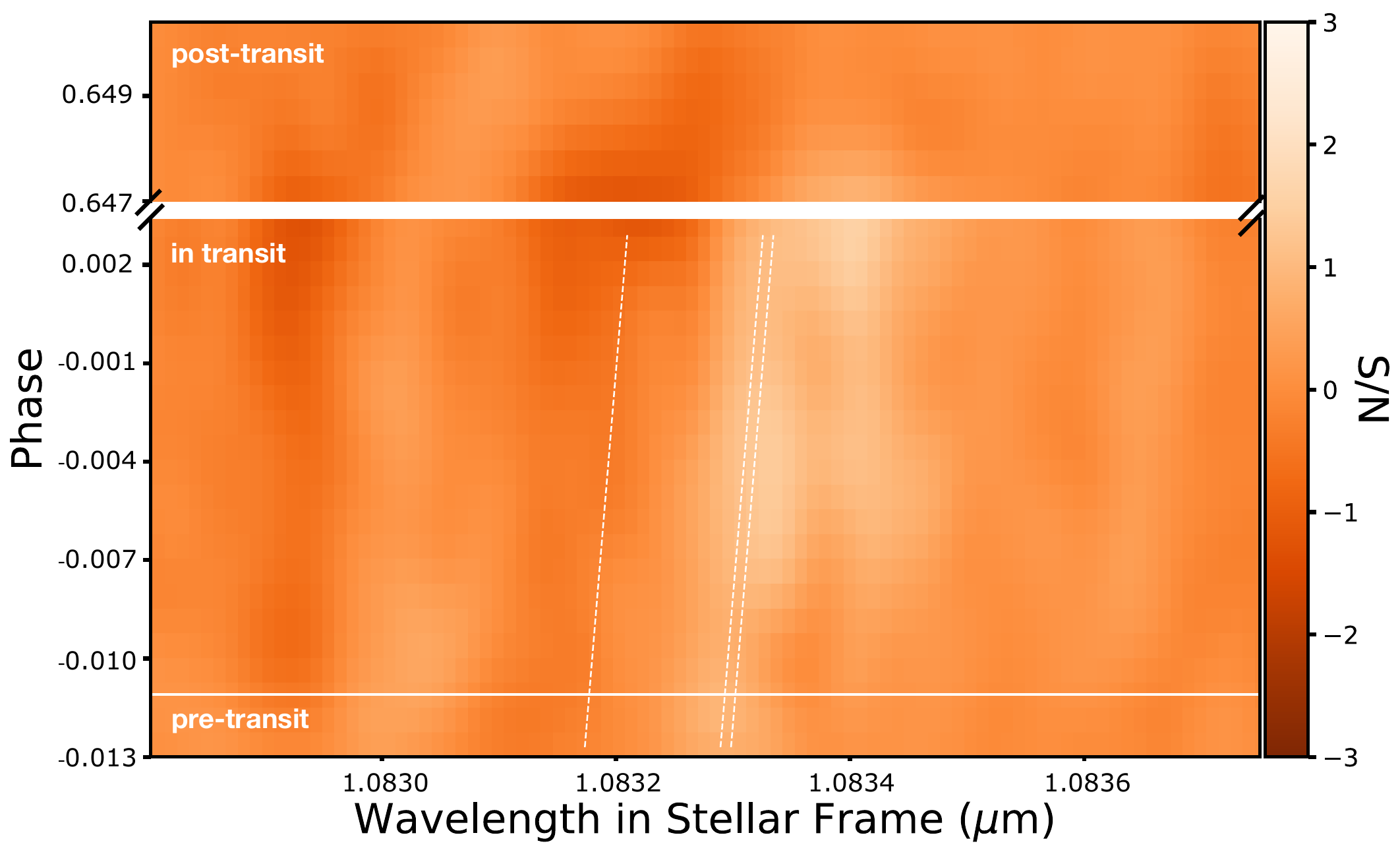}
    \caption{Phase-resolved signal-to-noise map for the partial transit of V1298 Tau c centered on the helium triplet in the stellar rest frame, compared to the first contact point (solid white lines) and the planet's orbital motion (dashed white lines).} \label{fig:v1298_He_absorption_map}
\end{figure*}

We first generated the mean in-transit and out-of-transit spectra in order to determine the in-transit excess absorption signal. We constructed the in-transit spectrum by computing the weighted mean (weighted by the spectral uncertainties) of spectra taken between the second and third contact points; the out-of-transit spectrum was constructed by computing the weighted mean of the spectra taken before first contact and after fourth contact. To determine the contact points, we used the ephemerides of V1298\,Tau\,c, HD 63433b, and K2-100b from \citet{David19}, \citet{Mann20}, and \citet{Barragan19}, respectively. The individual order 70 spectra, as well as the mean in- and out-of-transit spectra, for the planets in our sample are shown in Figures \ref{fig:k2100_it_oot}, \ref{fig:toi1726_it_oot}, and \ref{fig:v1298_it_oot}. We find that there is no apparent excess absorption centered on the helium triplet for any of the planets in our sample. We also constructed phase-resolved maps centered on the helium line in our search for excess absorption around 1.083 $\mu$m, computed by dividing each spectrum of the time-series by the mean out-of-transit spectrum. The phase-resolved excess absorption maps are shown in Figures \ref{fig:k2100_He_absorption_map}, \ref{fig:toi1726_He_absorption_map}, and \ref{fig:v1298_He_absorption_map}.  

We then constructed the helium transmission spectra for K2-100b, HD 63433b, and V1298\,Tau\,c, by shifting the excess absorption to each planet's respective rest frame, which are shown in Figure \ref{fig:He_tr_spec}. Since the native uncertainties in the transmission spectra are not representative of the scatter in the data due to the presence of correlated noise, we inflated our uncertainties to account for the actual scatter in the data. This step was necessary as, without it, we could infer anomalous or overly precise measurements of planetary absorption. We do this by setting the uncertainties in the helium transmission spectra to be equal to the standard deviation of the transmission spectra. From our spectra, we place 3$\sigma$ upper limits on the excess He absorption of 0.56\% for K2-100b, 0.47\% for HD 63433b and 1.13\% for V1298 Tau c. These upper limits are equal to three times the standard deviation of the (flat) transmission spectra taken from $\pm$10\,\AA\,around the He triplet. We also calculate upper limits on the equivalent width (EW) by integrating the transmission spectra between 10831--10835\,\AA. This leads to upper limits on the EW of $<$4.44\,m\AA\ for K2-100b, $<$2.52\,m\AA\ for HD 63433b and $<$8.49\,m\AA\ for V1298 Tau c.
% From our spectra, we place} \textbf{3$\sigma$} upper limits on the He excess absorption of 0.56\% (equivalent width, EW \textbf{$<$} 4.44\,m\AA), %$\pm$0.32\,m\AA), 0.47\% (EW \textbf{$<$} 2.52\,m\AA)%$\pm$0.28\,m\AA)
% , and 1.13\% (EW \textbf{$<$} 8.49\,m\AA) %$\pm$0.32\,m\AA) 
% for K2-100b, HD 63433b, and V2198 Tau c, respectively. \textbf{These upper limits are equal to three times} the standard deviation of the (flat) transmission spectrum taken from $\pm$10\,\AA\,around the He triplet.  
From both the mean in-transit and out-of-transit spectra as well as the excess absorption maps, we do not identify excess absorption for the planets in our sample. We use the {\tt p-winds} \citep{DosSantos22} code to place upper limits on the in-transit He absorption (\S \ref{sec:1d_models}).

%%%%%%%%%%%%%%%%%%%%%%%%%%%%%%%%%%%%%%%%%%%%%%%%%%%%%%%%%
\section{Helium Non-Detections in Context}
\label{sec:interpretation}
%%%%%%%%%%%%%%%%%%%%%%%%%%%%%%%%%%%%%%%%%%%%%%%%%%%%%%%%%

We consider our non-detections of excess helium absorption in light of the effects of stellar winds and magnetic fields on depopulating the metastable helium line (\S \ref{sec:conf_b_fields}), 1D isothermal escape models (Parker winds; \S \ref{sec:1d_models}), and photoevaporation-driven hydrodynamic escape models (\S \ref{sec:hydro}). 

%%%%%%%%%%%%%%%%%%%%%%%%%%%%%%%%%%%%%%%%%%%%%%%%%%%%%%%%%
\subsection{Stellar Winds \& Magnetic Fields}
\label{sec:conf_b_fields}
%%%%%%%%%%%%%%%%%%%%%%%%%%%%%%%%%%%%%%%%%%%%%%%%%%%%%%%%%
Given our non-detections, we should assess whether they agree with the basic predictions of outflow models. A hydrodynamically launched outflow can only be accelerated up to trans-sonic velocities. Beyond the sonic regime, the velocity remains approximately constant \citep[e.g.,][]{Lammerbook} and the density profile falls off as $1/r^2$. When the helium triplet state is depopulated by electron collisions (rather than photoionizations), as is expected for K-stars \citep{Oklopcic19}, the helium triplet fraction is roughly constant at large radius. This means that the helium absorption is dominated by the contribution from large radius, rather than close to the planet. Therefore, the helium absorption signal is dominated by how far the outflow extends spherically away from the planet. In the case of limited stellar wind, the outflow can ultimately only remain spherical out to the Hill sphere; whereas a strong stellar wind can crush the planetary outflow to a fraction of its Hill sphere \citep{Carolan2020}. Given a maximum radius out to which the outflow remains spherical ($R_{\rm out}$), the maximum possible excess absorption is:
\begin{equation}
\approx \left[1-\left(\frac{R_p}{R_{\rm out}}\right)^2\right]\left(\frac{R_{\rm out}}{R_*}\right)^2 \label{eqn:maximal}
\end{equation}
in the case that the outflow is optically thick in transmission at $R_{\rm out}$. Setting $R_{\rm out}$ to the Hill sphere radius $R_{\rm H}$ yields maximum excess optical depths of 1.9\%, 8.4\%, and 4.7\% for K2-100b, HD 63433b, and V1298 Tau c, respectively. The fact that these maximal excess depths are only a factor of $\sim$4 larger than our upper limits for K2-100b and V1298 Tau c implies that a transmission optical depth in helium of $\lesssim 0.25$ would be sufficient to render the helium absorption below our detection limit. Thus, since the helium line is normally optically thin, these non-detections are not surprising. For an optically thin $1/r^2$ density profile, the excess absorption becomes:
\begin{equation}
    \approx 2\sigma_{\rm He3} \pi n_{He3}(R_{\rm out})R_{\rm out} \left(1-\frac{R_p}{R_{\rm out}}\right)\left(\frac{R_{\rm out}}{R_*}\right)^2 \label{eqn:simple2}
\end{equation}
where $\sigma_{\rm He3}$ is the cross-section at 10830\,\AA~ (appropriately averaged over our R$\sim$25,000 spectral resolution) and $n_{He3}(R_{\rm out})$ is the density of metastable helium at $R_{\rm out}$. Even with the optimistic assumption that all helium is neutral and an extreme helium triplet fraction of $10^{-5}$, for a Solar ratio of hydrogen to helium the excess helium absorption remains below our upper limits for K2-100b and V1298 Tau c with mass-loss rates up to $10^{11}$ and $3\times10^{11}$~g~s$^{-1}$, respectively. These limits are consistent with typical energy-limited mass-loss rates. 

Applying the same methodology to HD 63433b would require a mass-loss rate below $10^{10}$~g~s$^{-1}$ due to its larger Hill sphere -- but this estimate is in tension with model expectations, of several $10^{10}$~g~s$^{-1}$. One immediate solution would be to reduce the outer radial extent of the outflow by its interaction with the stellar wind. To crush the planetary outflow to a sufficiently small size to be consistent with our non-detection would require a stellar wind with a velocity at the planet of several hundred km~s$^{-1}$ to have an outflow rate of $\sim$$10^{-12}~$M$_\odot$~yr$^{-1}$, approximately a hundred times larger than that of the Sun. Given the young age of the HD 63433 system and its observed X-ray properties, such a large stellar wind outflow rate is consistent with empirical scaling relationships \citep[e.g.][]{Wood2005,Johnstone2015,Blackman2016}. We note that a study using Zeeman-Doppler Imaging infers stellar mass-loss rates in the range $1-4 \times 10^{-13}$~M$_\odot$\,yr$^{-1}$ \citep{Bellotti24}.

An alternative explanation for HD 63433b is the presence of a strong planetary magnetic field. \citet{Schreyer23} demonstrated that any planetary magnetic field can suppress the helium transmission signal. This arises from two interrelated effects. Firstly, since the majority of the gas is confined by closed planetary magnetic fields, it cannot expand and cool as a result of stellar XUV heating. Thus, the gas is heated to $\sim$$10^4$~K where Lyman-$\alpha$ cooling becomes efficient, thermostating the gas at $\sim$$10^4$~K \citep[e.g.,][]{MurrayClay2009,Owen2016}. Around K-stars, the helium triplet state is depopulated by collisions with electrons \citep{Oklopcic19} and populated by recombinations -- processes that both happen less efficiently at high temperatures, resulting in a significantly lower population of metastable helium \citep{Biassoni2023}.  

Secondly, the outflowing regions near the pole have lower densities than a spherically symmetric outflow. This is because the outflow -- forced to follow planetary magnetic field lines -- has a streamline divergence that falls faster than $1/r^2$. In the outflow regions near the pole, the streamline divergence falls off as $1/r^3$ \citep{Adams2011,Owen2014}. This difference is important because for a $1/r^2$ density profile the helium absorption signal is dominated by absorption far from the planet, where it covers a larger area of the star (Equation~\ref{eqn:simple2}); whereas for $1/r^3$ the absorption is more weighted toward the regions closer to the planet where they cover less of the star's area. As \citet{Schreyer23} demonstrated, this effect significantly reduces the helium absorption for a planet with a strong magnetic field, where the excess absorption for a $1/r^3$, compared to a $1/r^2$ density profile is of order $R_p/R_{\rm out}$ smaller, which is $\sim$0.1 for HD 63433b. As discussed by \citet{Zhang_2022_HD63433_upper_limit_non_detect}, planetary magnetic fields could be responsible for the He non-detection. Following the methodology in \citet{Owen2014}, we estimate a surface dipole planetary magnetic field strength of $\sim$4-5~G to close $>$50\% of the planetary field lines.

%%%%%%%%%%%%%%%%%%%%%%%%%%%%%%%%%%%%%%%%%%%%%%%%%%%%%%%%%
\subsection{Parker Wind Models} 
\label{sec:1d_models}
%%%%%%%%%%%%%%%%%%%%%%%%%%%%%%%%%%%%%%%%%%%%%%%%%%%%%%%%%

We modeled the helium transmission spectra presented in \S \ref{sec:analysis} for HD 63433b, K2-100b, and V1298\,Tau\,c using the one-dimensional atmospheric escape model {\tt p-winds}\footnote{\url{https://github.com/ladsantos/p-winds}} \citep{DosSantos22,pwinds}. This open-source Python-based code is based on the framework of \citet{Oklopcic18} and \citet{Lampon20}, and has been benchmarked against the EVaporating Exoplanets (EVE) code (\citealt{Bourrier13,Bourrier15}). The model treats the escaping material as an isothermal Parker wind \citep{Parker58} composed only of hydrogen and helium, finds the steady-state recombination/ionization solutions for the distribution of neutral hydrogen and helium in the planetary upper atmosphere, and solves the radiative transfer equation to determine the in-transit absorption due to the escaping planetary material. The \texttt{p-winds} code requires as input high-energy spectral energy distributions (SEDs) of the host stars HD 63433, K2-100, and V1298\,Tau. The SEDs of V1298\,Tau and HD 63433 that we use are adapted from \citet{Duvvuri_2023_high_energy_sepc_of_v1298_tau} and \citet{Zhang22c}, respectively. For K2-100, since there is no archival high-energy spectrum available, we adopted the SED of the similarly young (600~Myr) and solar-type star $\iota$ Horologii \citep{SForcada2019} scaled to the system's stellar radius and semi-major axis. For these models, we assume that the outflow is composed solely of H and He, and that H/He ratio is 90/10 in number of particles. We adopt the stellar and planetary parameters in Table \ref{tab:params}.

Correlated noise is prevalent in the excess absorption spectra shown in Figure \ref{fig:He_tr_spec}.  Potential sources of correlated noise in the data may be instrumental or astrophysical in origin, including not limited to telluric contamination, wavelength calibration, stellar activity, and temporal variability in the the deep Mg 10811\,\AA, Si 10827\,\AA, and Si 10844\,\AA\,lines (e.g., \citealt{Guilluy23,Zhang23,Vissapragada24}). Therefore, we model the excess absorption spectra as a planetary absorption combined with correlated noise. To accurately constrain the mass loss, outflow temperature, and outflow velocity, we employ the use of Gaussian processes (GPs) to account for systematic uncertainties and model the correlated noise \citep{rasmussen2006gaussian}. The Matérn-3/2 kernel is used as the covariance function for the GPs, allowing for smoothness in our correlated noise functions \citep{foremanmackey17}. Gaussian processes allow for the fitting of the true helium absorption signal, removing the correlated noise present in the data. The use of Gaussian processes introduces two parameters (the systematics amplitude $\sigma$ and length scale $\rho$), which, when paired with the three \texttt{p-winds} parameters (mass-loss rate, outflow temperature, and outflow radial velocity), creates a five-dimensional model. The systematics and planetary absorption are fitted simultaneously. The Bayesian inference we designed avoids overfitting by applying a prior for the GP parameters inferred from the transmission spectrum outside of the He feature. This methodology will be discussed in more detail in an upcoming paper (McCreery et al., in prep.). When constraining the posteriors of these five parameters given the transmission spectra, we use nested sampling \citep{Speagle2020}. Nested sampling is advantageous over Monte Carlo methods in this situation due to our poor prior information and nested sampling's ability to estimate the marginal likelihood, which aids in model selection. 

\begin{figure*}
    \centering
    \includegraphics[width=\textwidth]{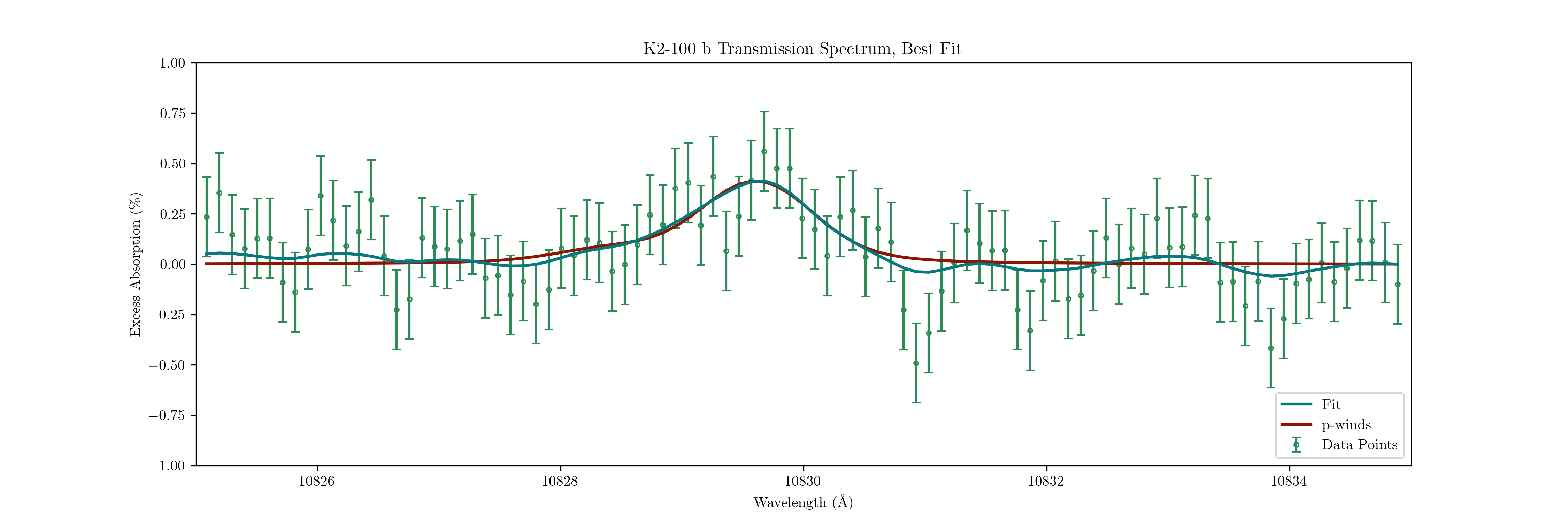}
    \includegraphics[width=\textwidth]{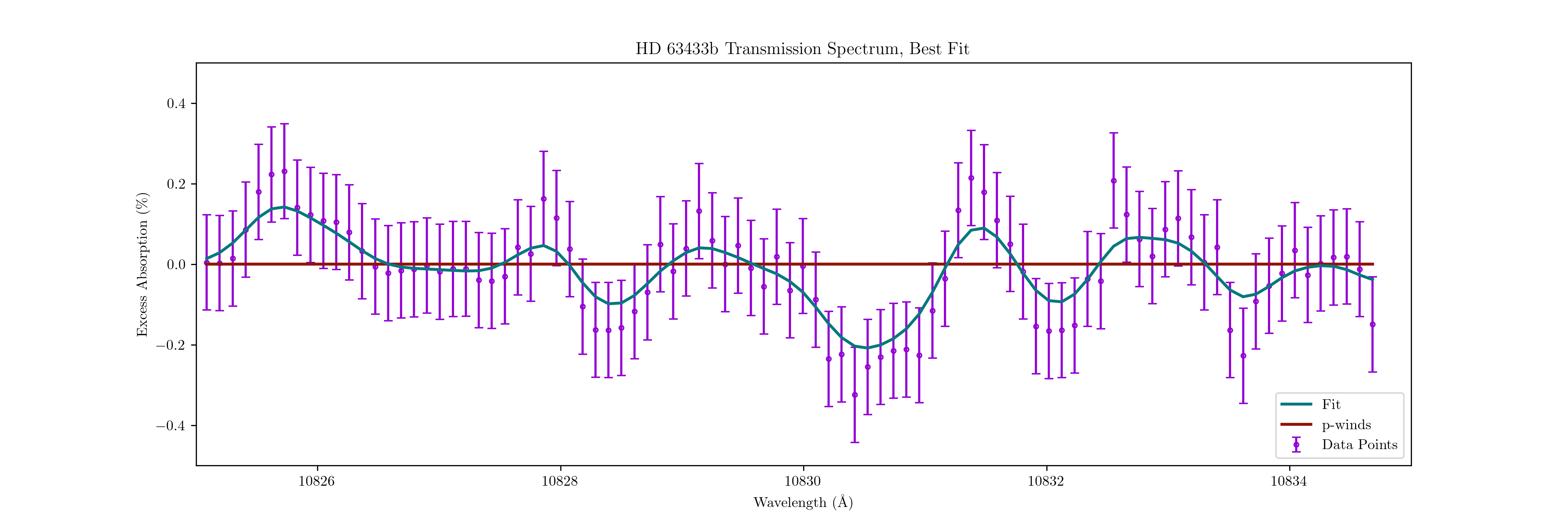}
    \includegraphics[width=\textwidth]{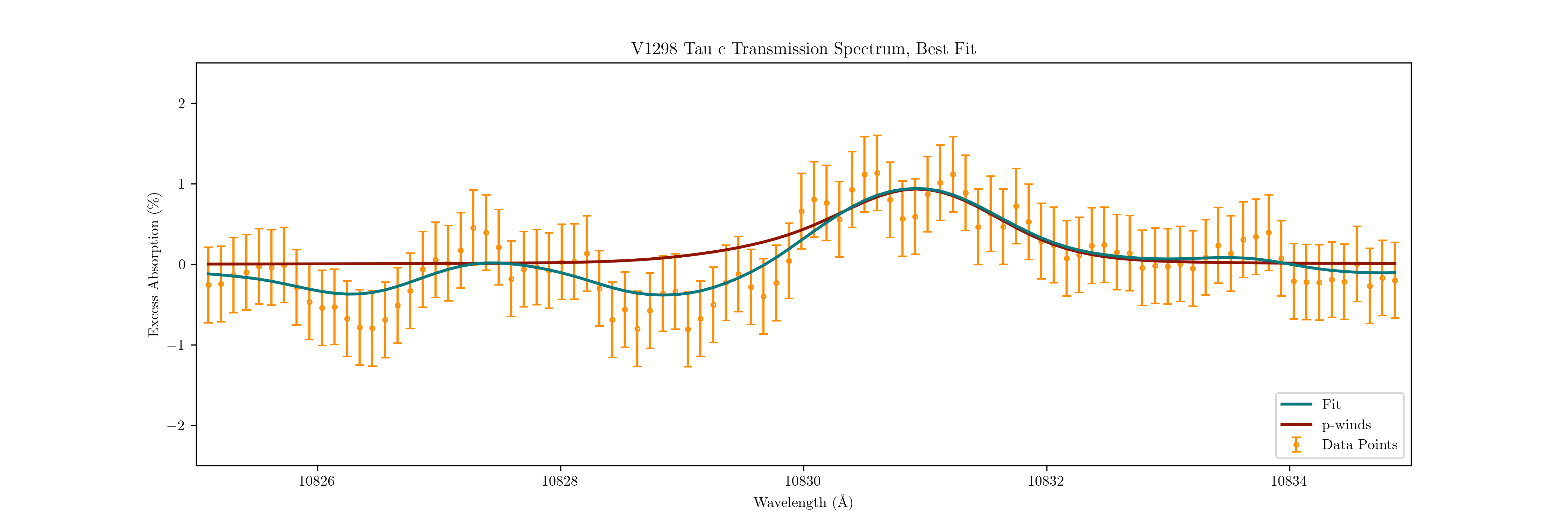}
    \caption{Best-fit results for simultaneously modeling the instrumental systematics and atmospheric escape signal.}
    \label{fig:sys_pwinds_fit}
\end{figure*}

In summary, our model uses nested sampling to perform fits of the helium absorption spectrum of HD 63433b, K2-100b, and V1298\,Tau\,c using \texttt{p-winds} to fit for the helium signature and Gaussian processes to account for correlated noise. The resulting best-fit systematics and planetary signals are shown in Figure \ref{fig:sys_pwinds_fit}. We found that the nested sampling results favor a non-detection for HD 63433b with more than 95\% confidence, a tentative detection at high blue-shifted velocities (-17~km\,s$^{-1}$) for K2-100b, and a marginal detection at high red-shifted velocities (+21~km\,s$^{-1}$) for V1298\,Tau\,c. To confirm that our model is not overfitting the data, we also ran fits only including the GP component. Table \ref{tab:logz_values} compares the results for the {\tt p-winds}+GP models to GP only models. We find similar posterior distributions from this exercise compared to the {\tt p-winds}+GP fits for V1298 Tau\,c and HD\,63433b. For K2-100b, the {\tt p-winds}+GP models are favored at the $\sim$3$\sigma$ level because -- although the length-scale of the GP is similar to the {\tt p-winds} feature, the amplitude is clearly larger at the location of the He line. 

\begin{table}
    \centering
    \begin{tabular}{lccc}
               & K2-100b & V1298 Tau c & HD 63433b \\
         \hline
         $\log z_{\text{GP}}$ & 461.17 & 392.43 & 490.59 \\
         $\log z_{\text{GP + } \texttt{p-winds}}$ & 464.53 & 391.57 & 489.17 \\
         $|\Delta \log z|, |\log B_m|$ & 3.36 & .85 & 1.42 \\
         $p$-value & .03 & .70 & .81\\
         \hline
    \end{tabular}
    \caption{Model comparisons between the Gaussian Process-only models and the full Gaussian Process + \texttt{p-winds} models using log-evidence ($\log z$). The Bayes factor, $B_m$, indicates that there are no significant differences in the model fits to the transmission spectra for V1298 Tau c and TOI-1726b. For K2-100b, however, the GP + \texttt{p-winds} model is moderately favored over the GP-only model.}
    \label{tab:logz_values}
\end{table}

\begin{figure*}
    \centering
    \includegraphics[width=\textwidth]{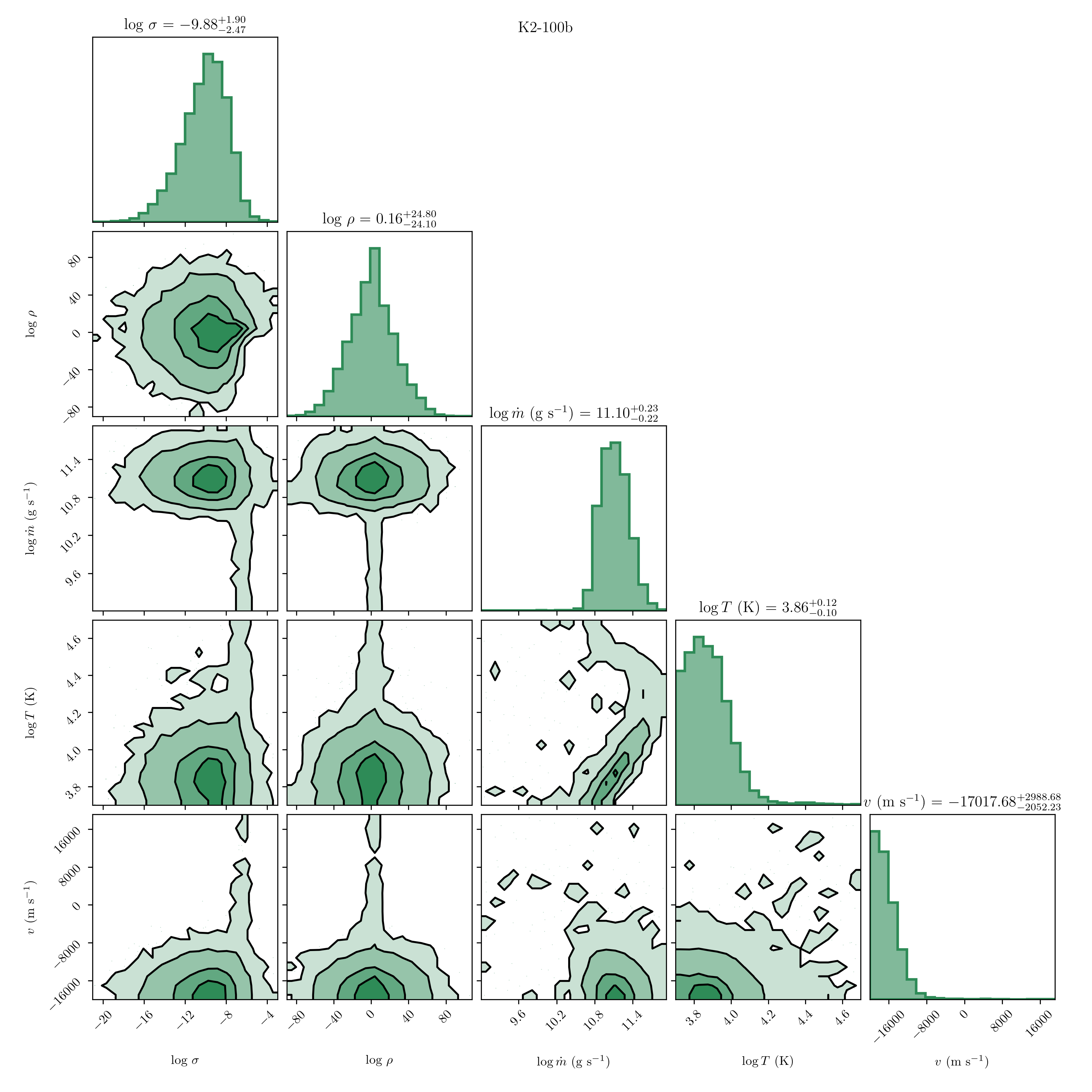}
    \caption{Posterior distributions of retrieved parameters from our 1D {\tt p-winds} models.}
    \label{fig:K2100b_corner}
\end{figure*}

\begin{figure*}
    \centering
    \includegraphics[width=\textwidth]{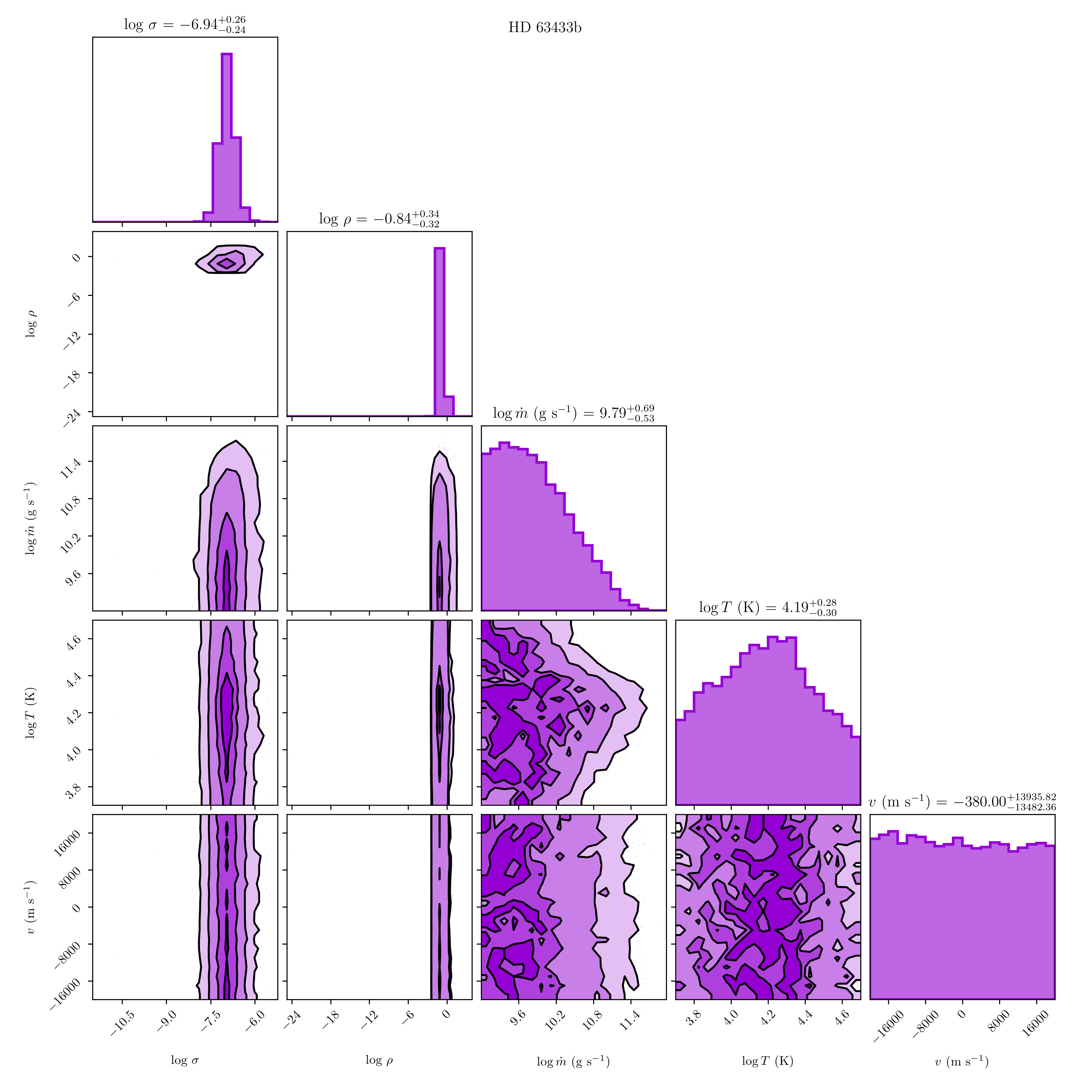}
    \caption{Same as Figure \ref{fig:K2100b_corner}, but for HD 63433b.}
    \label{fig:HD63433b_corner}
\end{figure*}

\begin{figure*}
    \centering
    \includegraphics[width=\textwidth]{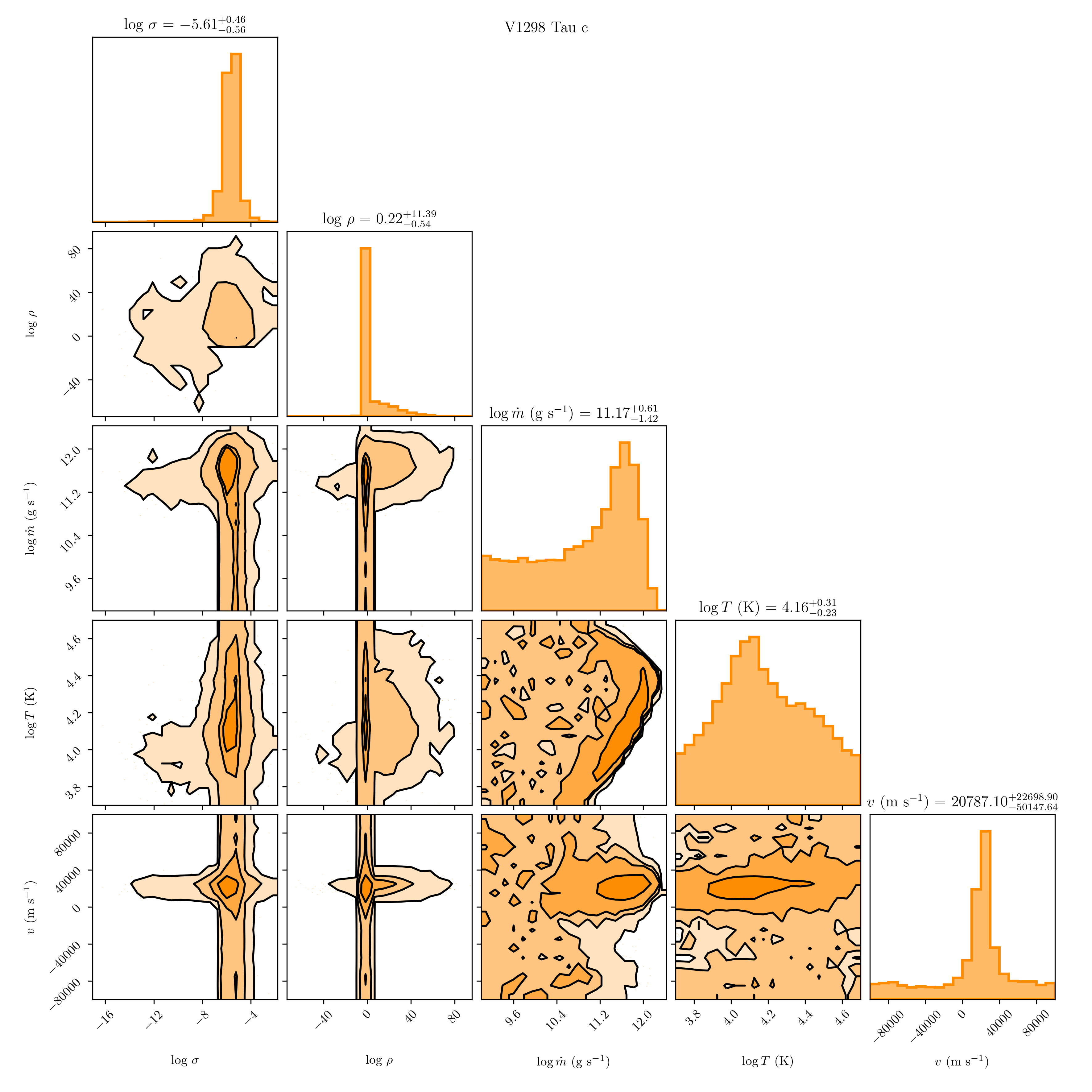}
    \caption{Same as Figure \ref{fig:K2100b_corner}, but for V1298 Tau c.}
    \label{fig:V1298Tauc_corner}
\end{figure*}

Most detections of helium outflows in transiting exoplanets have been found to exhibit blue-shifted velocities up to $\sim$$-10$~km\,s$^{-1}$ (McCreery et al.,\,in prep.), a result that is theoretically expected for confined winds \citep{MacLeod2022, Nail2024}. Slightly red-shifted helium signals are expected for outflows in magnetized exoplanets \citep{Schreyer23}. The outflow velocities measured by the nested sampling have never been observed before, are not predicted by state-of-the-art hydrodynamic escape models, and are likely spurious signals. The posterior distributions of the retrieved parameters from our {\tt p-winds} modeling are shown in Table \ref{tab:1d_models} and Figures \ref{fig:K2100b_corner}, \ref{fig:HD63433b_corner}, and \ref{fig:V1298Tauc_corner}. 

\begin{table*}
    \centering
    \caption{Parker-wind model constraints estimated for the observed transmission spectra}
    \label{tab:1d_models}
    \begin{tabular}{lcccc}
        Planet & Excess Absorption & log $\dot{m}$ (g\,s$^{-1}$) & T ($\times 10^3$ K) & v (km\,s$^{-1}$) \\
        \hline
        HD 63433b &  $<$2.52\,m\AA%$\pm$0.28\,m\AA 
        &  $<10.9$ (95\% conf.) & $>5000$ (prior) & No constraint \\
        V1298 Tau c & $<$8.49\,m\AA %$\pm$0.32\,m\AA 
        &  $<11.9$ (95\% conf.) & $>5000$ (prior) & $+21^{+23}_{-50}$ \\
        K2-100b & $<$4.44\,m\AA%$\pm$0.32\,m\AA 
        & $11.10^{+0.23}_{-0.22}$ & $7.2^{+2.3}_{-1.5}$ & $-17.0^{+3.0}_{-2.0}$ \\
        \hline
    \end{tabular}
\end{table*}

%%%%%%%%%%%%%%%%%%%%%%%%%%%%%%%%%%%%%%%%%%%%%%%%%%%%%%%%%
\subsection{Photoevaporation-driven Hydrodynamic H/He Models}
\label{sec:hydro}
%%%%%%%%%%%%%%%%%%%%%%%%%%%%%%%%%%%%%%%%%%%%%%%%%%%%%%%%%

% \input{Tables/tab_flux_params.tex}

%   Description of modelling 
Additionally, we modeled  HD 63433b, K2-100b, and V1298\,Tau\,c using the one-dimensional hydrodynamic model described in \citet{Allan23}. This model self-consistently solves the hydrodynamic equations of atmospheric escape simultaneously with equations tracking the state of hydrogen, as well as helium in its $1^1S$, $2^1S$, $2^3S$, singly and doubly ionized states. It considers the following hydrogen and helium processes: recombination, radiative decay, charge exchange between hydrogen and helium particles, collisional excitation, collisional ionization and photoionization. Fluxes in specified X-ray, soft and hard EUV, and mid-UV wavelength bins (see Table \ref{tab:params}) are required as input in order to model photoionizations. To this end, we use the same SEDs described in \S \ref{sec:1d_models}. Table \ref{tab:params} lists the resulting band integrated fluxes adopted in our modeling. Having modeled the hydrodynamics, we then utilize the ray tracing model also described in \citet{Allan23} to perform helium triplet transmission spectroscopy. 

\begin{figure*}
	\includegraphics[width=0.85\textwidth]{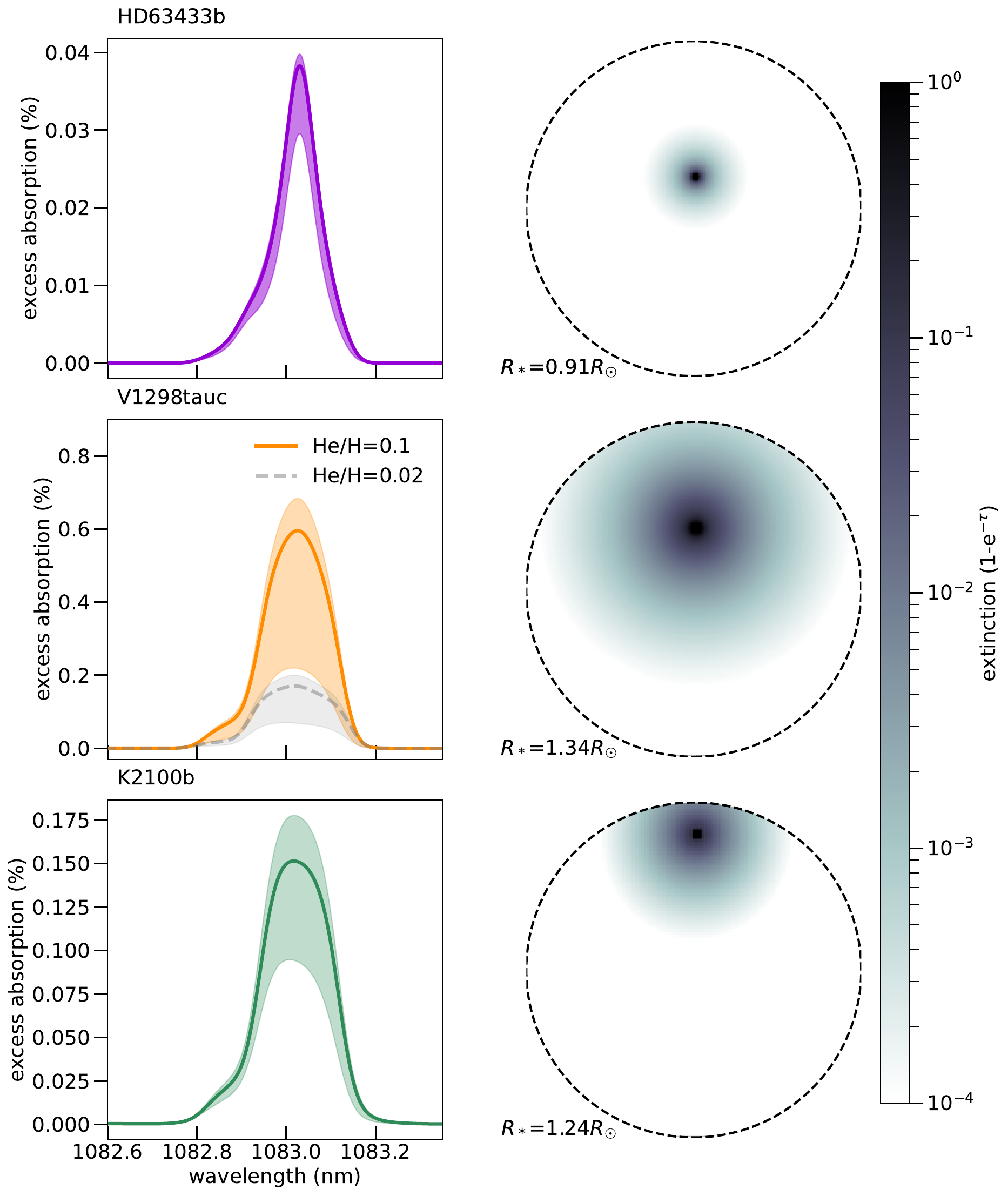}
    \caption{Helium triplet excess absorption predictions (left) 
    and their corresponding mid-transit extinction maps (right), obtained with the model of \citet{Allan23} as described in \S \ref{sec:hydro}. For the excess absorption profiles, solid lines indicate mean phase averaged (T1-T4) profiles while shaded region encompasses all phases between first contact and mid-transit. For the extinction maps, the dashed circle marks the stellar disk. The displayed extinction is the sum of the three individual extinctions of each of the triplet lines. A helium to hydrogen number fraction of He/H=0.1 was assumed for each planet, with He/H=0.02 also being tested in the case of V1298\,Tau\,c as shown by the gray dashed profile. Additional planetary and stellar parameters listed in Table \ref{tab:params} were used.} \label{fig:aa_predicted_abs}
\end{figure*}

Predicted helium triplet absorptions for our three planets using the \citet{Allan23} model are shown on the left-hand side of Figure \ref{fig:aa_predicted_abs}. The models have been convolved to the spectral resolution (R$\sim$25,000) of our observations. The solid lines show the mean average of phases between the transit's first and fourth points of contact, while the shaded region encompasses all phases between first-contact and mid-transit. The right-hand side of Figure \ref{fig:aa_predicted_abs} displays each planet's helium triplet extinction map across the stellar disk, as marked by the dashed circle. The displayed extinction is the sum of the individual extinction of each of the three lines of the triplet.

Despite assuming a large helium to hydrogen number fraction of He/H=0.1, the model predicts very weak absorption ($<$0.05\%) for HD 63433b, in agreement with the observations presented in Section \S \ref{sec:analysis}. Its low mass-loss rate prediction of 2.8 $\times 10^{10}$ g s$^{-1}$ is responsible for its low density and ultimately its weak helium triplet extinction, relative to the other planets.  

The model for V1298\,Tau\,c predicts a larger absorption more in line with the literature's current sample of detections with a mean predicted excess of 0.6\% assuming He/H=0.1, although we note that this prediction is still below the upper limit of 1.13\% we set from our observations. The predicted excess drops to 0.2\% if we instead assume He/H=0.02 in the hydrodynamic model, as shown by the gray dashed line. Accordingly, re-observing V1298\,Tau\,c with tighter uncertainties could help constrain the fraction of helium in this planet's atmosphere. Of the three planets, we predict V1298\,Tau\,c to have the highest mass-loss rate, 2.2 $\times 10^{11}$
g s$^{-1}$. The greater atmospheric escape rate is due to a favorable combination of high EUV flux and low surface gravity \citep{Allan_Vidotto_2019}. The large escape rate, combined with a mid-UV flux sufficiently low so as to prevent significant depopulation of helium triplet by mid-UV photoionization, are responsible for its greater He transit absorption predicted in our modeling.

The predicted phase averaged excess absorption for K2-100b is $\sim$0.15\%, also below the 3$\sigma$ upper limit of 0.56\% we get from the observation. Despite the low excess absorption, a large mass-loss rate of 1.1 $\times 10^{11}$ g s$^{-1}$ is predicted by the model.  The right-hand side of Figure \ref{fig:aa_predicted_abs} reveals that the weak absorption is partially due to its large impact parameter of 0.79$\rm R_\star$. 

\section{Discussion}
\label{sec:discussion}
%%%%%%%%%%%%%%%%%%%%%%%%%%%%%%%%%%%%%%%%%%%%%%%%%%%%%%%%%

%%%%%%%%%%%%%%%%%%%%%%%%%%%%%%%%%%%%%%%%%%%%%%%%%%%%%%%%%
\subsection{Comparison to Previous Observations}
%%%%%%%%%%%%%%%%%%%%%%%%%%%%%%%%%%%%%%%%%%%%%%%%%%%%%%%%%

%%%%%%%%%%%%%%%%%%%%%%%%%%%%%%%%%%%%%%%%%%%%%%%%%%%%%%%%%
\subsubsection{K2-100b}
%%%%%%%%%%%%%%%%%%%%%%%%%%%%%%%%%%%%%%%%%%%%%%%%%%%%%%%%%
High-resolution short-cadence spectroscopy from Subaru/IRD for K2-100b \citep{Gaidos20} yielded an upper limit on the planet's helium absorption of 5.7\,m\AA~in equivalent width and mass loss of $5.7 \times 10^{10}$\,g\,s$^{-1}$ (0.3\,$M_{\oplus}$\,Gyr$^{-1}$) using a Parker wind model with temperatures below $10^4$~K. With the Keck II/NIRSPEC data presented in this work, we can place tighter upper limits of $<$4.44
%$\pm$0.32
\,m\AA~ (0.56\%) excess helium absorption. From our He transmission spectrum, we measure a mass-loss rate of $1.25 \times 10^{11}\,\rm g\,s^{-1}$ from our Parker wind models (see \S \ref{sec:1d_models}), which is $\sim$5$\times$ higher than that of \citet{Gaidos20} and consistent with the \S \ref{sec:hydro} models. However, as we discussed in \S \ref{sec:1d_models}, our 1D Parker wind models for K2-100b converge to fit a likely spurious bump in our transmission spectrum and therefore our mass-loss measurements from this model are unreliable. Our theoretical 1D hydrodynamic models (\S \ref{sec:hydro}) predict a mass-loss rate of 1.1$\times 10^{11}$ g s$^{-1}$, which would correspond to He absorption of 0.15\% and is below our detection threshold (0.56\%). \cite{Gaidos20} have a tighter upper limit on the mass-loss than we obtain from our 1D hydrodynamic models, despite having a looser upper limit on the helium equivalent width.  This suggests that the differences in the mass-loss predictions from their study and ours may originate from the different model flux inputs, which would influence the resulting inferred transit absorptions. %due to a difference in the dynamics and metastable helium population. %in their use of 1D Parker wind models and our use of 1D hydrodynamic models (since our 1D Parker wind models are unreliable for K2-100b as mentioned previously). 

%FROM GAIDOS+ ABSTRACT: Our non-detection (< 5.7m ̊A) of a transit-associated He I line limits mass loss of a solar-composition atmosphere through a T ≤ 10000K wind to < 0.3 M⊕ Gyr−1. Either K2-100b is an exceptional desert-dwelling planet, or its mass loss is occurring at a lower rate over a longer interval, consistent with a core accretion-powered scenario for escape.  https://arxiv.org/pdf/2003.12940.pdf

%The predicted phase averaged excess absorption for K2-100b is $\sim$0.3\%, also below the upper limit of 0.56\% we get from the observation. Figure \ref{fig:aa_extinction} reveals that this is partially due to its large impact parameter of 0.79$\rm R_\star$. 

%%%%%%%%%%%%%%%%%%%%%%%%%%%%%%%%%%%%%%%%%%%%%%%%%%%%%%%%%
\subsubsection{HD 63433b}
%%%%%%%%%%%%%%%%%%%%%%%%%%%%%%%%%%%%%%%%%%%%%%%%%%%%%%%%%

\begin{figure*}
    \centering
    \includegraphics[width=\textwidth]{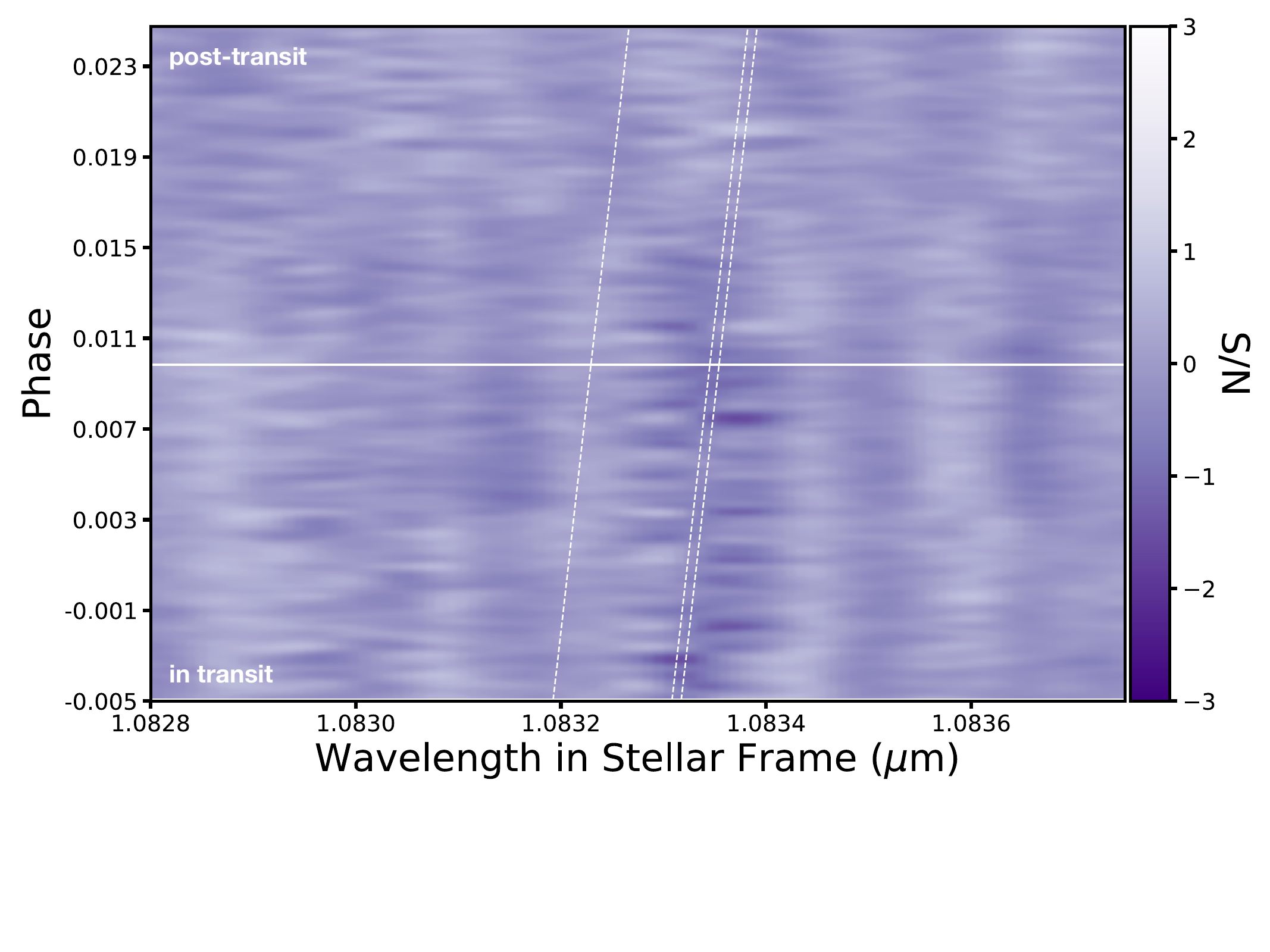}
    \caption{Phase-resolved signal-to-noise map of absorption for the partial transit (no pre-ingress baseline) of HD 63433b centered on the helium triplet in the stellar rest frame, compared to fourth contact (solid white line) and the planet's orbital motion (dashed white lines). }
    \label{fig:toi1726_He_absorption_map}
\end{figure*}

Previous Keck II/NIRSPEC observations of the HD 63433 system \citep{Zhang22c} placed an upper limit on HD 63433b's helium absorption of 0.5\,\%, which is consistent with the 0.47\,\% upper limit that we derive in this work. \cite{Zhang22c} also noted that the stellar helium line varied during their transit observations, demonstrated by a darkening in the helium absorption map during the transit of HD 63433b. This darkening matches what we see in the data presented here (Figure \ref{fig:toi1726_He_absorption_map}), and induces an apparent helium \textit{emission} feature (negative excess absorption) in the planet's transmission spectrum (Figure \ref{fig:He_tr_spec}). Similar to \cite{Zhang22c}, we attribute this apparent emission to a darkening in the stellar helium triplet that is unrelated to the planet's atmosphere, suggesting that the stellar helium triplet is variable over the timescale of a transit observation. \cite{Zhang22c} observed HD 63433b on 2021 January 7, two weeks after our observations. If this variability was to be seen in further subsequent epochs to have a period commensurate with the planet's orbital period, this could point to star-planet interactions, as has been proposed to explain the observed variation in the helium triplet for AU Mic \citep{Klein2022}.% The repeated nature of this variability may be related to planet-star interactions, although further observations are needed to test this hypothesis. 

%\cite{Zhang22c} also presented an upper limit on Lyman-$\alpha$ absorption for HD\,63433b. For HD\,63433c, they detected Lyman-$\alpha$ absorption of $11.5 \pm 1.5$\,\% and placed an upper limit on helium absorption of 0.5\,\%.
Using 3D hydrodynamic mass-loss models, \cite{Zhang22c} compute a mass-loss rate of $6.6 \times 10^{10}$\,g\,s$^{-1}$ (0.35\,$M_{\oplus}$\,Gyr$^{-1}$) for HD\,63433b, which implies a mass-loss timescale of 80\,Myr. This timescale is consistent with the 95th percentile upper limit on the mass loss rate of $7.9 \times 10^{10}\,\rm g\,s^{-1}$ (0.42\,$M_{\oplus}$\,Gyr$^{-1}$) derived from our Parker wind models, but higher than the rate $2.8 \times 10^{10}$\,g\,s$^{-1}$ (0.15 \,$M_{\oplus}$\,Gyr$^{-1}$) inferred from our self-consistent hydrodynamic models (\S \ref{sec:1d_models} and \ref{sec:hydro}), respectively. Given the age of HD 63433b ($414 \pm 23$\,Myr; \citealt{Mann20}), our {\tt p-winds} and self-consistent model findings agree with \cite{Zhang22c} in demonstrating that it is unlikely that the planet has retained a H/He atmosphere. 

Our 1D hydrodynamic modeling yields a mass-loss rate that is $\sim$2.3 times smaller than the \citet{Zhang22c} estimates, which consequently results in a smaller overall estimated outflow density. Although our model uses the same high-energy stellar SED as \citet{Zhang22c}, their model differs from ours in their treatment of the heating/cooling processes of the atmosphere. They also include the interaction with a stellar wind, but we believe the former difference is the likely cause of our differing evaporation rates. Because of the smaller escape rates and densities of our spherically symmetric models, it is expected that our excess helium absorption would differ from theirs: in fact, our excess helium absorption at the line center ($<$0.05\%) is an order of magnitude smaller than the model from \citet{Zhang22c} for HD 63433b. 

%We note, however, that our 1D hydrodynamic modeling yields a mass-loss rate that is $\sim$2--3 times smaller than the \citet{Zhang22c} estimates, which consequently results in a smaller overall estimated outflow density. Our 1D models provide a maximum of 0.06\% excess helium absorption, whereas the \citet{Zhang22c} excess absorption for HD 63433b is a factor of 2 larger. These differences arise because \citet{Zhang22c} use a 3D hydrodynamic model which incorporate the effects of a stellar wind that interacts with the planet at around 100$R_{\oplus}$ ($\approx$50\,$R_{p}$), whereas our 1D models do not incorporate stellar winds and are only computed out to $\sim$10\,$R_{p}$. These model differences account for the factor of 2 discrepancy in absorption and differing escape rates that we note. 
 
%%%%%%%%%%%%%%%%%%%%%%%%%%%%%%%%%%%%%%%%%%%%%%%%%%%%%%%%%
\subsubsection{V1298\,Tau\,c}
%%%%%%%%%%%%%%%%%%%%%%%%%%%%%%%%%%%%%%%%%%%%%%%%%%%%%%%%%

\begin{figure*}
    \centering
    \includegraphics[width=\textwidth]{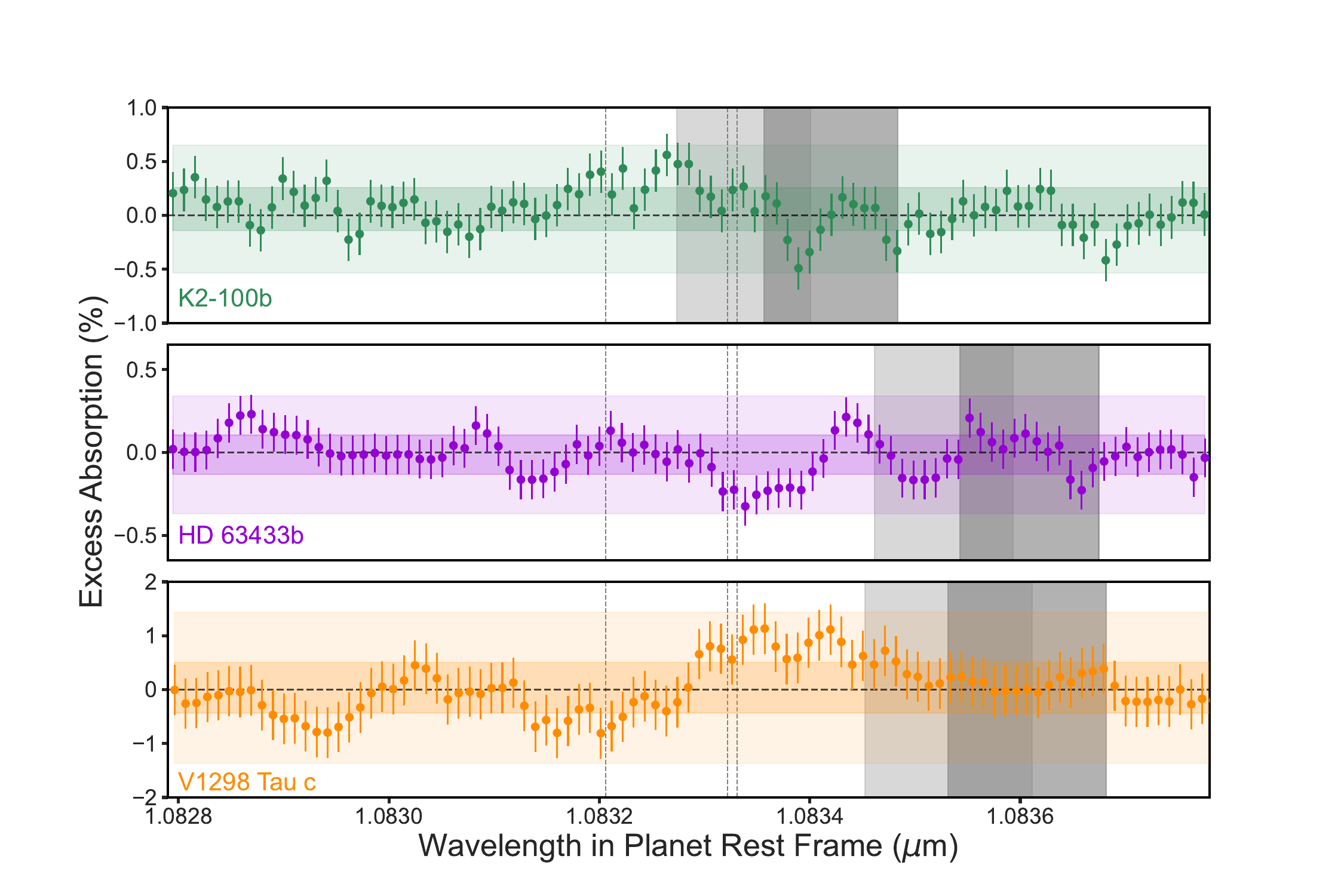}
    \caption{The He transmission spectrum for K2-100b (top), HD 63433b (middle), V1298 Tau c (bottom) centered on the helium triplet (dashed gray lines). The dark and light shaded colored regions correspond to the 1$\sigma$ and 3$\sigma$ uncertainties, respectively. Telluric regions are shown in the shaded gray regions, as in Figures \ref{fig:k2100_it_oot}, \ref{fig:toi1726_it_oot}, and \ref{fig:v1298_it_oot}.} 
    \label{fig:He_tr_spec}
\end{figure*}

\citet{Vissapragada21} used narrowband photometry from Palomar/WIRC to search for helium in V1298\,Tau\,c, but were unable to detect the planet's transit likely due to correlated noise in the data. Their non-detection is consistent with our findings for this planet (Figure \ref{fig:He_tr_spec}), however, our upper limit on the excess helium absorption of 1.13\% (EW $<$ 8.49
% $\pm$0.08
\,m\AA) is the first quantitative upper limit for this planet. We checked whether our result was sensitive to our choice of ephemeris by re-doing our analysis with the ephemeris of \cite{Feinstein22} and found a consistent upper limit to when we used the ephemeris from \cite{David19}. High-resolution optical spectra from GRACES on Gemini-North \citep{Feinstein21} revealed a tomographic signal in the Ca II triplet attributed to possible star-planet interactions, as well as excess H$\alpha$ absorption smoothly decreasing during the planet's transit that can be explained by starspots and faculae rotating into and out of view. 

Since the Ca II triplet arises in the chromosphere, searching for variability in other chromospheric lines may corroborate the star-planet interactions hypothesis \citep{Feinstein21}. Given that the helium triplet is also a chromospheric line, we can search for variability in this line \citep{Klein22}. However, the variability we see in the stellar helium line, combined with the poor weather conditions during our observations, prevents us from testing the possibility of star-planet interactions.

% the time-series stellar spectra for V1298\,Tau\,c shown in Figure \ref{fig:v1298_it_oot} demonstrate that the noise in our observations would prevent us from determining whether we are seeing true stellar variability as opposed to noise from these observations taken in poor weather conditions. %the helium light curves for V1298\,Tau\,c over the course of our observations. We see evidence of stellar helium variability, which is consistent with the star-planet interactions suggested by \cite{Feinstein21}. However, 
% Additional observations would be beneficial to further test for star-planet interactions.

% Beyond V1298\,Tau\,c, \cite{Vissapragada21} reported an upper limit on helium absorption in V1298\,Tau\,b's atmosphere of $R_b/R_{\star} < 0.019$, consistent with \cite{Gaidos22}'s non-detection of helium for V1298\,Tau\,b using Subaru/IRD. \cite{Vissapragada21} also reported tentative evidence for helium in the atmosphere of V1298\,Tau\,d of $\Delta R_d/R_{\star} = 0.0205 \pm 0.054$. However, this required a significant timing offset from the expected time of mid-transit. 

Variability in V1298 Tau's stellar helium line has been seen on a number of previous occasions \citep[e.g.,][]{Vissapragada21,Gaidos22,Krolikowski24}. \cite{Vissapragada21} measured a difference in the stellar helium equivalent width of 0.15\,\AA~over a two month timescale in addition to a flare in the helium line during a transit of V1298\,Tau\,c. \cite{Gaidos22} found a 50\,\% variability in the stellar helium line during a transit observation V1298\,Tau\,b. In this case, the variability is neither due to stellar variability due to the short time scale involved, nor planet variability given the orbital phases at which this absorption is seen. \cite{Gaidos22} instead suggest that the absorption they see could be associated with a transit of V1298\,Tau\,d, which would be consistent with \cite{Vissapragada21}. We rule out seeing variability in the stellar He line at the amplitudes and timescales that \citet{Gaidos22} see, reinforcing the conclusion that they were seeing absorption from V1298\,Tau\,d.   

V1298\,Tau was also observed as part of \cite{Krolikowski24}'s program to monitor the stellar He equivalent width of exoplanet host stars. For V1298\,Tau, the youngest star in their sample, they observed variability in the He equivalent width with a median absolute deviation of $61.5^{+8.4}_{-7.8}$\,m\AA\ or equivalently $3.93^{+0.54}_{-0.50}$\,\% in absorption depth over their observing baseline of $\sim$1.5 years. Over a three week time scale, the difference between our transit observations and subsequent additional out-of-transit baseline for V1298\,Tau, \cite{Krolikowski24} measure variability of $\sim$80\,m\AA, serving as a cautionary tale against using out of transit data from different epochs for these young systems. 

We also measure variability in the stellar helium line between our December and January out-of-transit observations. Over this 3 week time scale, the helium equivalent width varied from 453.784\,m\AA\ to 418.649\,m\AA\ (integrated between 10831.58\,\AA\ and 10835.48\,\AA). The equivalent widths and variability are similar to that seen by \cite{Krolikowski24}. We were motivated to use the January out-of-transit spectra because we were only able to acquire three out-of-transit spectra immediately before the December transit due to poor weather. However, in doing so, we are insensitive to planetary absorption because of the large amplitude variability in the stellar helium line. For this reason, we recalculated V1298\,Tau\,c's transmission spectrum using only the December out-of-transit data. This approach also reveals no in-transit He absorption with an upper limit of 0.55\,\%, which is consistent with our original approach of using the January data, and hence does not change our conclusions.

%%%%%%%%%%%%%%%%%%%%%%%%%%%%%%%%%%%%%%%%%%%%%%%%%%%%%%%%%
\subsection{Young Planets in Context}
%%%%%%%%%%%%%%%%%%%%%%%%%%%%%%%%%%%%%%%%%%%%%%%%%%%%%%%%%

\begin{figure*}
    \centering
    \includegraphics[width=\textwidth]{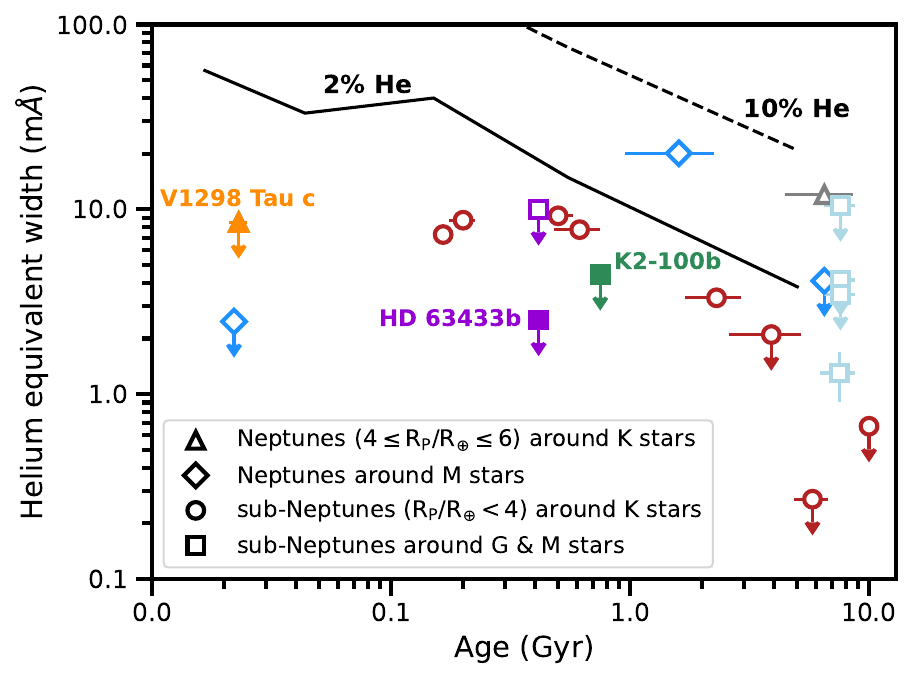}
    \caption{The measured equivalent width of the metastable helium line as a function of age for Neptunes (4 $\leq$ $\rm R_{\oplus}$ $\leq$ 6) around K stars (triangles) and M stars (diamonds), and sub-Neptunes ($\rm R_{\oplus} <$ 4) around K stars (circles) and around G and M stars (squares). The upper limits from our sample are shown for K2-100b (green), V1298\,Tau\,c (orange), and HD 63433b (purple). The He equivalent width for HD 63433b from \citet{Zhang22c} is shown in the open purple square. We also show predicted trends between helium absorption and age from hydrodynamic models \citep{Allan23} for 2\% (solid black line) and 10\% (dashed black line) He abundance. These models are calibrated for a 0.1\,M$\rm{_{Jup}}$ planet orbiting a K-type star at 0.045\,au (details in text). Literature helium values are taken from \protect\cite{Allart18,Gaidos23,Kasper20,Krishnamurthy21,Nortmann18,Orell-Miquel22,Orell-Miquel23,Zhang21,Zhang22d,Zhang23}.}
    \label{fig:models}
\end{figure*}

% AA: Having modelled more suitable smaller planets, and noticing EW value issue replacing this par. 
%     Analysis of models to observations will need to be redone after plot is updated with new EW evolution. 
% To contextualize the helium non-detections presented here for K2-100b, HD 63433b, and V1298 Tau c, we compare our derived He upper limits to measured helium equivalent widths from the literature for sub-Jovian ($\rm R_{\oplus}$ $\leq$ 6) planets orbiting K dwarfs and other stellar host types in Figure \ref{fig:models}. We compare these helium measurements to the trend from 1D hydrodynamic models  \citep{Allan23} of helium absorption as a function of system age, which demonstrate an exponential decay in He absorption with age for 2\% and 10\% He abundances. While Neptunes and sub-Neptunes planets around other host spectral types (squares and diamonds) do not follow the trend from \citet{Allan23}, sub-Neptunes around K dwarfs (circles) from the literature and our upper limit for V1298\,Tau\,c (a Neptune around a K star) match this trend for 2\% He abundance well in both detections and upper limits. The other two sub-Neptunes in our sample, which orbit G stars (K2-100b and HD\,63433b), also roughly fall along the 2\% He abundance. 
To contextualize the helium non-detections presented here for K2-100b, HD 63433b, and V1298 Tau c, we compare our derived He upper limits to measured helium equivalent widths from the literature for sub-Jovian ($\rm R_{\oplus}$ $\leq$ 6) planets orbiting K dwarfs and other stellar host types in Figure \ref{fig:models}. We also compare these helium measurements to 1D hydrodynamic modeling predictions with planetary evolution. We extend upon the work of \citet{Allan23}, by computing two new self-consistent model sets better suited to the three planets discussed in this work, namely a smaller 0.1\,$\rm M_{J}$ planet orbiting the same K-type star at 0.045\,au, again considering both 2\% and 10\% He abundances. For the radius evolution input, we use the models of \citet{Fortney_Nettelmann2010} for a 0.1\,$\rm M_{J}$ planet, with a 25\,$\rm M_{\oplus}$ planetary core. Over the considered evolution timeframe of 16 to 5000\,Myr, the radius evolves from 0.83 to 0.55\,$\rm R_{J}$, closer to -- but still above -- the radii of the planets here (0.346, 0.192, 0.499 and \,$\rm R_{J}$ for K2-100b, HD 63433b, and V1298 Tau c, respectively). Other than these smaller radius and mass inputs, the modeling of the evolution predictions presented here is identical to the F2 and F10 model sets of \citet{Allan23}. 

Compared to these previous larger, inflated models, the two new self-consistent evolution model sets experience a stronger gravitational force due to the smaller planetary radius (despite the lower planetary mass assumed). Consequently, the predicted atmospheric escape is also lower. During transit, this, in addition to the smaller planetary radii results in relatively reduced coverage of the stellar disk by atmospheric triplet-state helium and ultimately a lower predicted observability of escape. However, the overall trend of weaker He absorption with evolution noted in \citet{Allan23} remains in the new 0.1\,$\rm M_{J}$ model-sets, with the solid and dashed profiles of Figure \ref{fig:models} corresponding to 2\% and 10\% He abundances.

Considering the non-detections we find here and the high estimated mass-loss rates from both our hydrodynamic escape models (isothermal Parker wind and photoevaporation-driven escape), we consider the ages of the systems in our sample and the timescales of photoevaporation assuming a constant mass-loss rate with evolution. For HD\,63433\,b, the self-consistent model's predicted mass-loss rate of $2.8 \times 10^{10}$\,g\,s$^{-1}$ gives an evaporation timescale of 215 Myr if we adopt the same envelope fraction of 0.6\% \citep{Zhang22c}. This rough estimation is also likely underestimated due to our assumption of a constant mass-loss rate with evolution, due to the escape likely being stronger at ages younger than the sampled age in reality. Given HD\,63433\,b's predicted age (414$\pm$23 Myr; \citealt{Mann20}), both {\tt p-winds} and our hydrodynamic models are consistent with \citep{Zhang_2022_HD63433_upper_limit_non_detect} in demonstrating that it is unlikely that the planet has retained a H/He atmosphere. 

Applying the constant mass loss with evolution approximation to V1298\,Tau\,c and K2-100\,b results in photoevaporation timescales of 140 and 225 Myr, respectively. Using a more ample envelope fraction of 2\% for mini-Neptunes leads to timescales of 465 and 740 Myr. Regardless of the assumed envelope fraction, the very expected young age of V1298\,Tau\,c ($23\pm4$\,Myr) suggests that it should still have its primordial atmosphere. K2-100b -- with its age of $750 \pm 5$ Myr -- may be close to its photoevaporation lifetime, depending on the mass fraction of the envelope. Furthermore, the hydrodynamic modeling (which assumes a primordial atmosphere) shows that the loss of the primordial atmosphere is not required to explain the non-detections -- the low number density of helium triplet material obscuring the disk alone is enough to do so. 

% have since ran models more suitable to the smaller sizes of the obs. 
% It should be noted that the evolving hydrodynamic predictions of \citet{Allan23} were performed for a larger 0.3\,$\rm M_{J}$ planet orbiting a K-type star at 0.045\,au, with a radius evolving from 18.9 to 11.8\,$\rm R_{\oplus}$ as the system aged from 16 to 5000\,Myr \citep{Fortney_Nettelmann2010}. These larger radii will enhance the helium triplet absorption of the \citet{Allan23} evolving models in two main ways. The first being the greater obscuring area during transit across the stellar disk. Secondly, as it will cause a weaker gravitational force, they will undergo enhanced atmospheric escape and hence produce a higher number density of obscuring material. 

We therefore propose three different explanations for the He non-detections presented here: i) it is possible that the planets formed with a low He abundance; ii) stellar winds and/or magnetic field effects quash the He signal (\S \ref{sec:conf_b_fields}); or iii) escape is occurring, but helium is not in the metastable state (\S \ref{sec:hydro}) and thus not observable with our 1.083\,$\mu$m data. Magnetic fields are consistent with the HD 63433b non-detection, while the K2-100 and V1298 Tau c non-detections do not require magnetic fields and can instead be explained by a combination of the Hill radii and likely optically thin nature of He being less than our detection threshold, combined with some potential stellar wind confinement.    

% \textcolor{magenta}{From formation models, we do not expect the planets to have such a low He abundance. So it seems that either we are seeing observational evidence of fractionated escape (not hydrodynamic escape) or perhaps the self-consistent models are missing something (for instance, metals that cool the outflow).}

Intriguingly, the majority of previously published He {\sc i} observations for young sub-Jovian planets have yielded several non-detections (e.g., \citealt{Gaidos20,Vissapragada21,Zhang22c,Zhang23}) -- contrary to theoretical predictions of higher or ongoing atmospheric escape rates early on in the evolutionary timescales of these systems. Long-term monitoring of stellar helium variability for a sample of mostly young stars \citep{Krolikowski24} -- including the three planets that are the subject of this study -- show that exoplanet helium absorption is typically less than the stellar helium variability at ages $<300$ Myr. This %finding aligns with the signals that we see for the planets presented here, and 
presents another confounding factor for better understanding young planets.

%https://arxiv.org/pdf/2311.01313.pdf

%compare to other small planet He searches (https://arxiv.org/pdf/2209.03502.pdf)

% compare to helium pop \citep{Bennett23}, https://arxiv.org/pdf/2308.02002.pdf (Table 3 shows all detections and non-detections)

% compare to a young planet He detection (https://arxiv.org/pdf/2307.05191.pdf, https://arxiv.org/pdf/2307.09515.pdf, https://arxiv.org/pdf/2307.15024.pdf)

%%%%%%%%%%%%%%%%%%%%%%%%%%%%%%%%%%%%%%%%%%%%%%%%%%%%%%%%%
\section{Conclusions} 
\label{sec:summary}
%%%%%%%%%%%%%%%%%%%%%%%%%%%%%%%%%%%%%%%%%%%%%%%%%%%%%%%%%

Using Keck II/NIRSPEC, we observed the helium transmission spectra for three young ($<$800 Myr) sub-Jovian (0.2--0.5\,$\rm R_{J}$) planets: HD 63433b, K2-100b, and V1298 Tau c. We find upper limits on the excess helium absorption ranging from 0.47--1.13\% for all three planets in this sample, with upper limits on their mass-loss rates of $7.9\times10^{10} - 7.9\times10^{11}$\,g\,s$^{-1}$ from Parker wind models (\S \ref{sec:1d_models}). These observed upper limits are consistent with the analytical maximum excess absorption predictions at 1.083\,$\mu$m ($\sim$2--8\%; \S \ref{sec:conf_b_fields}) and estimates of $<$0.04--0.6\% from 1D hydrodynamic escape models (\S \ref{sec:hydro}). In the context of other He observations of sub-Jovian planets in both young and older systems, our results follow theoretical predictions for the exponential decay of helium absorption in sub-Jovians as a function of system age. %and are consistent with 1D hydrodynamic models calibrated to a starting He abundance of 2\,\%. 
Further 1.083\,$\mu$m observations for a larger sample of sub-Jovian planets at a range of ages is crucial for better understanding the temporal evolution of the helium triplet.

\section*{Acknowledgments}
% \begin{acknowledgments}
We thank Jorge Sanz-Forcada for kindly sharing the high-energy SED of $\iota$ Horologii. We acknowledge W. M. Keck Observatory, which is operated as a scientific partnership among the California Institute of Technology, the University of California and the National Aeronautics and Space Administration. The Observatory was made possible by the generous financial support of the W.M. Keck Foundation. The authors wish to recognize and acknowledge the very significant cultural role and reverence that the summit of Maunakea has always had within the indigenous Hawaiian community. We are most fortunate to have the opportunity to conduct observations from this mountain.
JK acknowledges financial support from Imperial College London through an Imperial College Research Fellowship grant. APA and AAV acknowledge funding from the European Research Council (ERC) under the European Union's Horizon 2020 research and innovation programme (grant agreement No 817540, ASTROFLOW). R. A. is a Trottier Postdoctoral Fellow and acknowledges support from the Trottier Family Foundation. This work was supported in part through a grant from the Fonds de Recherche du Québec - Nature et Technologies (FRQNT). This work was funded by the Institut Trottier de Recherche sur les Exoplanètes (iREx). This work has been carried out within the framework of the NCCR PlanetS supported by the Swiss National Science Foundation under grants 51NF40$\_$182901 and 51NF40$\_$205606. This project has received funding from the European Research Council (ERC) under the European Union's Horizon 2020 research and innovation programme (project {\sc Spice Dune}, grant agreement No 947634). 
% \end{acknowledgments}

\bibliography{main}
\bibliographystyle{aasjournal}

\end{document}